\documentclass[aps,prx,reprint,superscriptaddress,nobibnotes,longbibliography]{revtex4-2} 
\pdfoutput=1

\usepackage{graphicx}
\usepackage{amsmath,amssymb}
\usepackage{xspace}
\usepackage{siunitx}
\usepackage[
	pdftex,
	colorlinks=true,
  	urlcolor=blue,
  	linkcolor=blue,
  	citecolor=blue
]{hyperref}
\usepackage[caption=false]{subfig}
\usepackage{textcomp}

\newcommand{\ket}[1]{\ensuremath{\vert #1 \rangle}\xspace}%

\begin{document}

\title{Site-resolved imaging of ultracold fermions\\ in a triangular-lattice quantum gas microscope}

\author{Jin Yang}
\thanks{These authors contributed equally to this work.}
\author{Liyu Liu}
\thanks{These authors contributed equally to this work.}
\author{Jirayu Mongkolkiattichai}
\thanks{These authors contributed equally to this work.}
\author{Peter Schauss}
\affiliation{Department of Physics, University of Virginia, Charlottesville, Virginia 22904, USA}
\email{ps@virginia.edu}

\begin{abstract}
Quantum gas microscopes have expanded the capabilities of quantum simulation of Hubbard models by enabling the study of spatial spin and density correlations in square lattices. However, quantum gas microscopes have not been realized for fermionic atoms in frustrated geometries. Here, we demonstrate the single-atom resolved imaging of ultracold fermionic $^{6}$Li atoms in a triangular optical lattice with a lattice constant of \SI{1003}{\nano\meter}. The optical lattice is formed by a recycled narrow-linewidth, high-power laser combined with a light sheet to allow for Raman sideband cooling on the $D_1$ line. We optically resolve single atoms on individual lattice sites using a high-resolution objective to collect scattered photons while cooling them close to the two-dimensional ground vibrational level in each lattice site. By reconstructing the lattice occupation, we measure an imaging fidelity of \mbox{$\sim98\%$}. Our new triangular lattice microscope platform for fermions clears the path for studying spin-spin correlations, entanglement and dynamics of geometrically frustrated Hubbard systems which are expected to exhibit exotic emergent phenomena including spin liquids and kinetic frustration.
\end{abstract}

\maketitle

\section{Introduction}
Frustrated quantum systems pose a significant challenge to condensed matter theory due to their extensive ground state degeneracy \cite{Wannier1950,Anderson1987} and can show fractional quasi-particle statistics as known from quantum Hall physics \cite{Wen1989}. There are a wide variety of interesting phenomena in frustrated systems. Examples include spin liquids, time-reversal symmetry breaking, and kinetic constraints \cite{Balents2010,Batista2016,Zhou2017}. While small systems can be solved with tremendous computational resources, predictions for the low-temperature phases in the thermodynamic limit are scarce and often debated \cite{Yoshioka2009,Shirakawa2017,Szasz2020}. Existing condensed matter realizations are complicated materials and simpler model systems are sought after. 
Ultracold atoms provide a unique way to explore quantum many-body physics through quantum simulations of frustrated quantum systems based on first principles. Prominent examples for quantum simulation with ultracold atoms include the direct detection of antiferromagnetic correlations \cite{Greif2013,Hart2015,Drewes2016b,Parsons2016,Boll2016,Cheuk2016,Brown2017,Gall2021} and the observation of many-body localization \cite{Gross2017}. Ultracold atoms in optical lattices implement Hubbard models \cite{Lewenstein2007,BlochDalibardZwerger2008,Esslinger2010}, where neighboring sites are coupled by hopping and atoms interact if they meet on the same lattice site. Fermi-Hubbard systems were first realized with ultracold atoms in square lattices \cite{Joerdens2008,Schneider2008}. 
Frustrated lattice geometries have been studied with absorption imaging of ultracold bosonic atoms \cite{Becker2010} which led to quantum simulation of classical frustration \cite{Struck2011}. Other geometrically frustrated two-dimensional lattice geometries like kagome lattices \cite{Jo2012} and the Lieb lattice \cite{Taie2015} have been studied with bosonic atoms, and recently individual bosonic atoms have been imaged in a triangular lattice \cite{Yamamoto2020}. 
But for the implementation antiferromagnetic interactions, fermions are the more natural choice \cite{Tieleman2013}. For revealing intricate correlations on short length scales, this asks for a fermionic quantum gas microscope, where all ultracold atoms in the many-body system can be imaged simultaneously. Existing fermionic quantum gas microscopes were used to study Hubbard models on square lattices \cite{Cheuk2015,Parsons2015,Haller2015,Edge2015,Omran2015,Greif2016,Brown2017}. However, to obtain a geometrically frustrated system a non-bipartite lattice geometry is required. The triangular lattice is the paradigm example of a frustrated lattice \cite{Wannier1950}, because a triangle is the simplest structure where antiferromagnetic constraints cannot be simultaneously satisfied on all bonds. For triangular lattices the frustration of antiferromagnetic order leads to a remarkable quantum phase transition for varying interaction between a magnetically ordered state and a disordered state which may be a chiral spin liquid \cite{Shirakawa2017,Szasz2020}.

Here, we demonstrate the first realization of a site-resolved quantum gas microscope of ultracold fermionic atoms in a triangular lattice, thereby paving the way for a new platform to study frustrated Hubbard physics in a lattice with spacing of 1003 nm and strong tunneling in the tight-binding limit. We load a degenerate Fermi gas into the triangular lattice and obtain densities above half filling.

Our experiment uses fermionic $^{6}$Li because it possesses intriguing properties like broad Feshbach resonances and a low mass, allowing to realize Hubbard models at larger tunneling and greater interaction than with other species. However, the low mass comes with difficulties localizing the atoms during imaging. Therefore, we rely on fluorescence imaging during Raman sideband cooling near the ground state in the lattice. Raman sideband cooling of lithium is challenging due to the large lattice depth required to suppress tunneling and to reach the Lamb-Dicke regime. The triangular lattice adds to this difficulty due to extensive required optical access and constraints on the beam geometry. We designed a sophisticated lattice setup to overcome these obstacles, allowing optically resolved imaging of individual fermionic atoms in the triangular optical lattice with high fidelity.

In this paper, we first discuss our experimental setup to prepare $^6$Li degenerate Fermi gases and our novel approach to create a triangular lattice for ultracold atoms. Then we present detailed information about the implementation of single-site resolved imaging in the triangular lattice via Raman sideband cooling. We discuss reconstruction of the lattice occupation from imaging results and the imaging fidelity in a comparative study of single-atom imaging in three different optical lattices. We conclude with an outlook on the study of Fermi-Hubbard physics and frustrated quantum physics in our setup.

\begin{figure*}
\centering
\includegraphics[width=\linewidth]{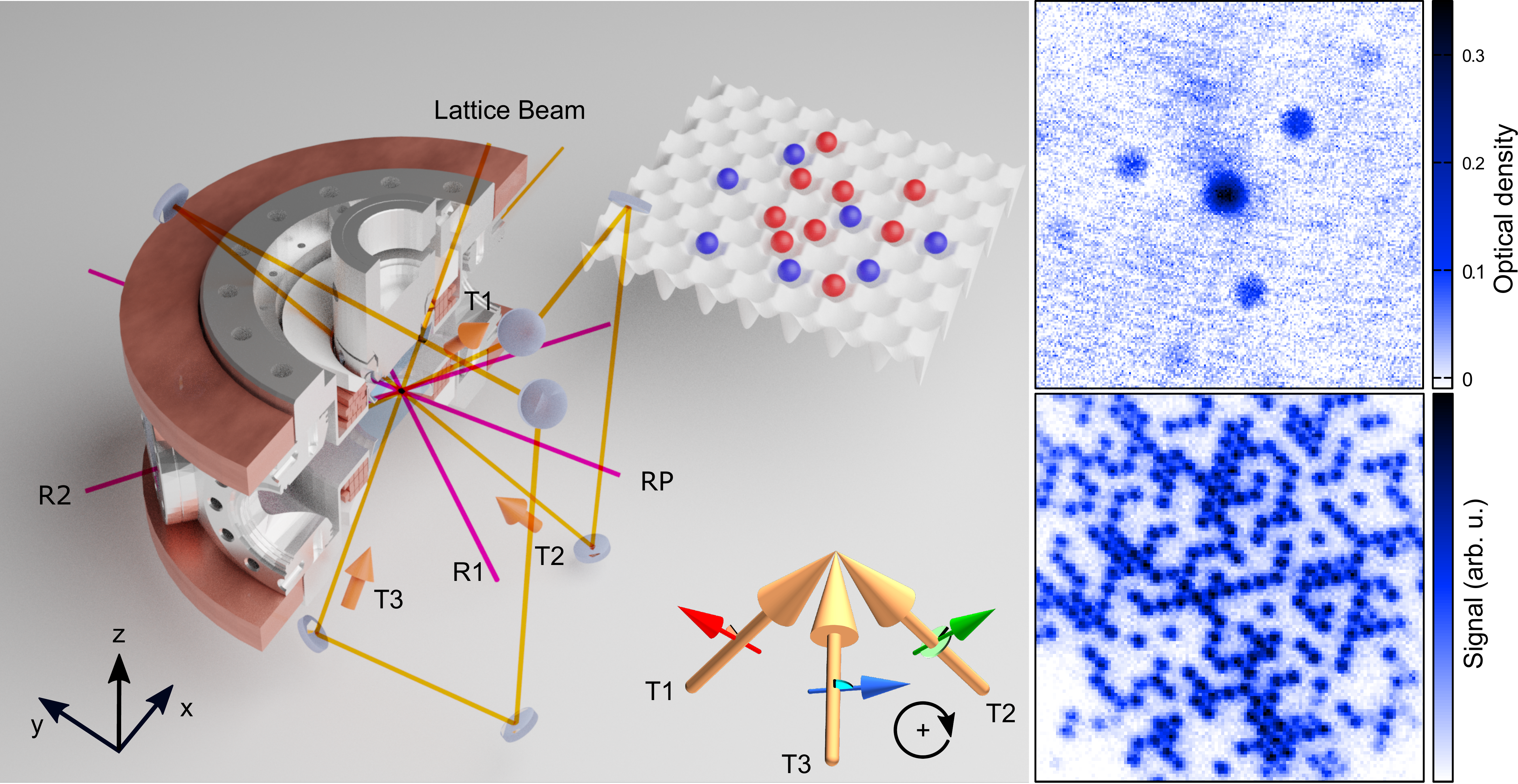}
\caption{{\bf Triangular-lattice quantum gas microscope.} (\emph{left}) Sketch of triangular lattice and Raman sideband imaging beams and their alignment relative to the vacuum chamber. The stainless steel octagon chamber is equipped with an outer copper coil pair for the MOT field and inner coil pair for the Feshbach field. The triangular lattice is formed by recycling the lattice beam through the recessed top and bottom windows, leaving just enough space for the objective at the top window. The second and third focus are created by 1\,:\,1 imaging systems, which are not shown. Three orange arrows (T1, T2 and T3) indicate the direction of the three beams which cross at the position of the atoms where the triangular lattice is formed. The polarization configuration used for imaging in the lattice is illustrated in the bottom middle inset. The Raman cooling beams (R1 and R2) and the Raman repump beam (RP) are sent through the side windows. (\emph{top right}) Kapitza-Dirac scattering of  $^{6}$Li molecular Bose-Einstein condensate (BEC) from the triangular lattice. This image is an average of 10 absorption images after a time-of-flight of \SI{1.5}{\milli\second}, using about 1\% of the maximum lattice laser power and a pulse length of \SI{2}{\micro\second}. For this picture, we used polarization angles of $0^\circ$ for lattice beams T1, T2, and T3 to demonstrate a symmetric lattice. (\emph{bottom right}) Raw site-resolved fluorescence image of $^{6}$Li atoms in the triangular lattice.}
\label{fig: Sketch of the lattice setup and Raman sideband imaging setup}
\end{figure*}

\section{Experimental Setup}

\subsection{Preparation of degenerate Fermi gas}
In the following, we describe the experimental setup and the path to a degenerate Fermi gas. For stability and fast cycle time of the experiment, we designed a single-chamber experiment that allows for sufficient optical access for all required laser beams [Fig.~\ref{fig: Sketch of the lattice setup and Raman sideband imaging setup}].
We start with 600,000 $^{6}$Li atoms in a $\sim \SI{1070}{\nano\meter}$ crossed optical dipole trap (CDT) loaded from a magneto-optical trap (MOT). The MOT is loaded via a Zeeman slower. To increase loading efficiency of the CDT, we use a compressed MOT stage where the power for both the cooling light and the repump light is decreased to 0.01\% and the detuning is changed from \SI{-30}{\mega\hertz} to \SI{-5}{\mega\hertz} within \SI{4}{\milli\second}. The CDT is formed by crossing an incoming laser beam (YLR-300-LP-AC-Y14) with its retroreflection under an angle of $\sim$ 10$^{\circ}$, where the power of each beam is \SI{125}{\watt} with a beam waist of \SI{90}{\micro\meter} at the crossing point. The atoms evenly populate the states $\ket{1}$ $\equiv$ $\ket{\textrm{2$^{2}$S$_{1/2}$ F = 1/2 m$_{F}$ = 1/2}}$ and $\ket{2}$ $\equiv$ $\ket{\textrm{2$^{2}$S$_{1/2}$ F = 1/2 m$_{F}$ = $-1/2$}}$ after loading into the CDT, and the initial density is $\sim$ \SI{1d12}{\centi\meter^{-3}} with a temperature of $\sim$ \SI{220}{\micro\kelvin}. Thereafter, a three-stage evaporation (plain evaporation for \SI{1}{\second}; forced evaporation I for \SI{0.7}{\second}; forced evaporation II for \SI{5}{\second}) leads to a $^{6}$Li degenerate Fermi gas. During the evaporation, a Feshbach field is ramped up to 810\,G, where the scattering length between state $\ket{1}$ and $\ket{2}$ is $a_{s}\approx17,000a_{0}$ \cite{Bartenstein2005, Zuern2013}, where $a_0$ is the Bohr radius. The intensity of the dipole trap stays unchanged in plain evaporation. In forced evaporation I, the intensity of the dipole trap is reduced to 6\% of the initial value following an exponential decay curve with a time constant $\tau_{1}$ = \SI{300}{\milli\second}. In forced evaporation II, the intensity of the dipole trap continues to reduce to 0.4\% of the initial value with a time constant $\tau_{2}$ = \SI{6}{\second}. To prevent the formation of deeply bound lithium molecules and obtain degenerate Fermi gases, the Feshbach magnetic field is switched from 810\,G to 300\,G ($a_{s} = -288 a_{0}$) within \SI{10}{\milli\second} and about \SI{500}{\milli\second} before the end of the forced evaporation stage II, where the density is not yet high enough to form lithium dimers via three-body collisions. With this experimental cycle of \SI{12}{\second} duration, we obtain a degenerate Fermi gas with about 3,000 atoms and temperature below one fifth of the Fermi temperature, determined by a Fermi fit to a non-interacting gas.

\subsection{Triangular lattice setup}
Fluorescence imaging of atoms in the triangular geometry requires a strong three-dimensional confinement at each lattice site. Therefore, we need to find a triangular lattice configuration that provides sufficient lattice depth at the limited available laser power. For this purpose, we interfere three laser beams to create a triangular array of one-dimensional light tubes and add a strongly oblate ``light sheet" beam to complete the three-dimensional confinement. The strongly oblate light sheet has beam waists of $\SI{4.2}{\micro\meter}\times\SI{50}{\micro\meter}\times\SI{70}{\micro\meter}$ and uses power of \SI{24}{\watt} at \SI{1070}{\nano\meter}. The trap frequency along $z$ axis is $\sim$ \SI{160}{\kilo\Hz}. In order to create a deep triangular lattice with resolvable lattice spacing we use an unusual approach. We recycle a single \SI{1064}{\nano\meter} laser beam (MOPA 55W, Nd:YAG, Coherent) twice and cross all three beams at the position of the light sheet, thereby reusing the laser power three times [Fig.~\ref{fig: Sketch of the lattice setup and Raman sideband imaging setup}]. The phases of the three lattice beams do not need to be stabilized because phase drifts only lead to translations of the triangular lattice. To keep these translations within tolerable bounds of about one lattice site per minute, the setup is very rigid and temperature-controlled via water cooling and air conditioning. 

All three lattice beams propagate from the negative $z$ direction (down) to the positive $z$ direction (up) with an angle of 45$^{\circ}$ out of the $x$-$y$ plane. Their projections onto $x$-$y$ plane cross to each other at an angle of 120.0(6)$^{\circ}$. The power for each beam is \SI{42}{\watt}, \SI{40}{\watt} and \SI{38}{\watt}, respectively, due to losses caused by optics during the recycling. All three beams have a Gaussian beam waist of $\sim$ \SI{30}{\micro\meter} at the crossing. This leads us to a triangular lattice with a lattice spacing of $a_\text{latt}$ = \SI{1003}{\nano\meter}. Our configuration for the lattice is compatible with a standard octagon vacuum chamber but requires very careful consideration of objective mount and magnetic field coils which typically block the optical access exploited here, as illustrated in Fig.~\ref{fig: Sketch of the lattice setup and Raman sideband imaging setup}. In addition, we have a custom-designed anti-reflection coating for the vacuum windows to reduce the reflection at the 45$^{\circ}$ angle of incidence. 
Since the interference pattern between the three crossing beams depends both on the wavevector direction and the polarization of each beam, these parameters have to be carefully adjusted for each beam. The angles between the lattice beams are restricted to about $1^\circ$ by the optical access and we use half-wave plates to control the polarizations of all lattice passes. For the following experiments, we adjusted these to obtain the strongest possible interference pattern in the triangular lattice. We found that the lattice depth is maximal for incoming linear polarization angles of about $40^\circ, -40^\circ$, and $80^\circ$ for lattice beams L1, L2, and L3, respectively, relative to the vertical polarization closest aligned to the $z$ axis [Fig.~\ref{fig: Sketch of the lattice setup and Raman sideband imaging setup}]. 
Due to birefringence in the vacuum windows and coatings the polarizations may be slightly modified at the atom position. The asymmetry of the configuration leads to anisotropic tunneling in the lattice in our current configuration. We confirmed by explicit calculation that anisotropic triangular lattice geometries can be adiabatically transformed to a symmetric configuration by varying the polarization of one of the three lattice beams. To implement such a scheme, we plan to add the capability to dynamically switch between the maximum-lattice-depth and an isotropic-tunneling configuration during the experimental cycle by upgrading to a motorized wave plate mount in the future. 
 
To prepare a quasi-two-dimensional Fermi gas in the triangular lattice, we first load the degenerate Fermi gas from the CDT into the light sheet and evaporate for another \SI{250}{\milli\second}. The intensity of the light sheet is reduced to 0.2\% of its initial value following an exponential decay curve with a time constant $\tau_3$ = \SI{100}{\milli\second}. This evaporation is necessary to remove excitations created during the loading procedure. Next, the intensity of the light sheet is increased to the initial value again, and the triangular lattice is adiabatically switched on within \SI{100}{\milli\second}. This configuration with maximal depth of lattice and light sheet is used for imaging the atoms by collecting fluorescence during Raman sideband cooling.
 
For calibration of the lattice depth, we carried out Kapitza-Dirac scattering [analogous to Fig.~\ref{fig: Sketch of the lattice setup and Raman sideband imaging setup}] and measured the atom number in the zeroth order as a function of lattice intensity. Through fitting of the decay curve to a Bessel function, we find a maximum lattice depth of \mbox{$\sim5000\,E_{r}$} with $E_{r} \equiv \hbar^2\pi^2/(2ma_{\text{latt}}^2)=\SI{8.2}{\kilo\Hz}$.

\section{Raman sideband cooling}
In order to keep the atoms localized at each single site during the fluorescence imaging, we utilize Raman sideband cooling to collect scattered photons while keeping the atoms near the ground-state of the harmonic potential. Variations of Raman sideband cooling have been used to detect various atomic species in optical lattices with single-atom resolution \cite{Li2012,Haller2015,Cheuk2015,Parsons2015,Edge2015,Omran2015}. A two-photon Raman sideband transition transfers atoms from one hyperfine ground state to the other hyperfine ground state while lowering the vibrational level in the on-site harmonic trap. The frequency difference between the two photons needs to be calibrated to match the frequency difference between the two hyperfine ground states plus the lattice on-site harmonic oscillator frequency $\omega_\text{latt}$. To close the cooling cycle, the atoms need to be transfered back to the initial hyperfine ground state without changing their vibrational levels. This is implemented through an optical pumping process using the Raman repump laser. In order to keep the heating in the repump process low, a large $\omega_\text{latt}$  is required to suppress recoil heating in $x$-$y$ plane by operating in the Lamb-Dicke regime. After many cycles, most atoms occupy the ground vibrational level which is a dark state in the absence of heating processes. The scattered photons in the optical pumping process are then collected to image the atoms. 

\begin{figure}[ht]
	\centering
	\includegraphics[width=\linewidth]{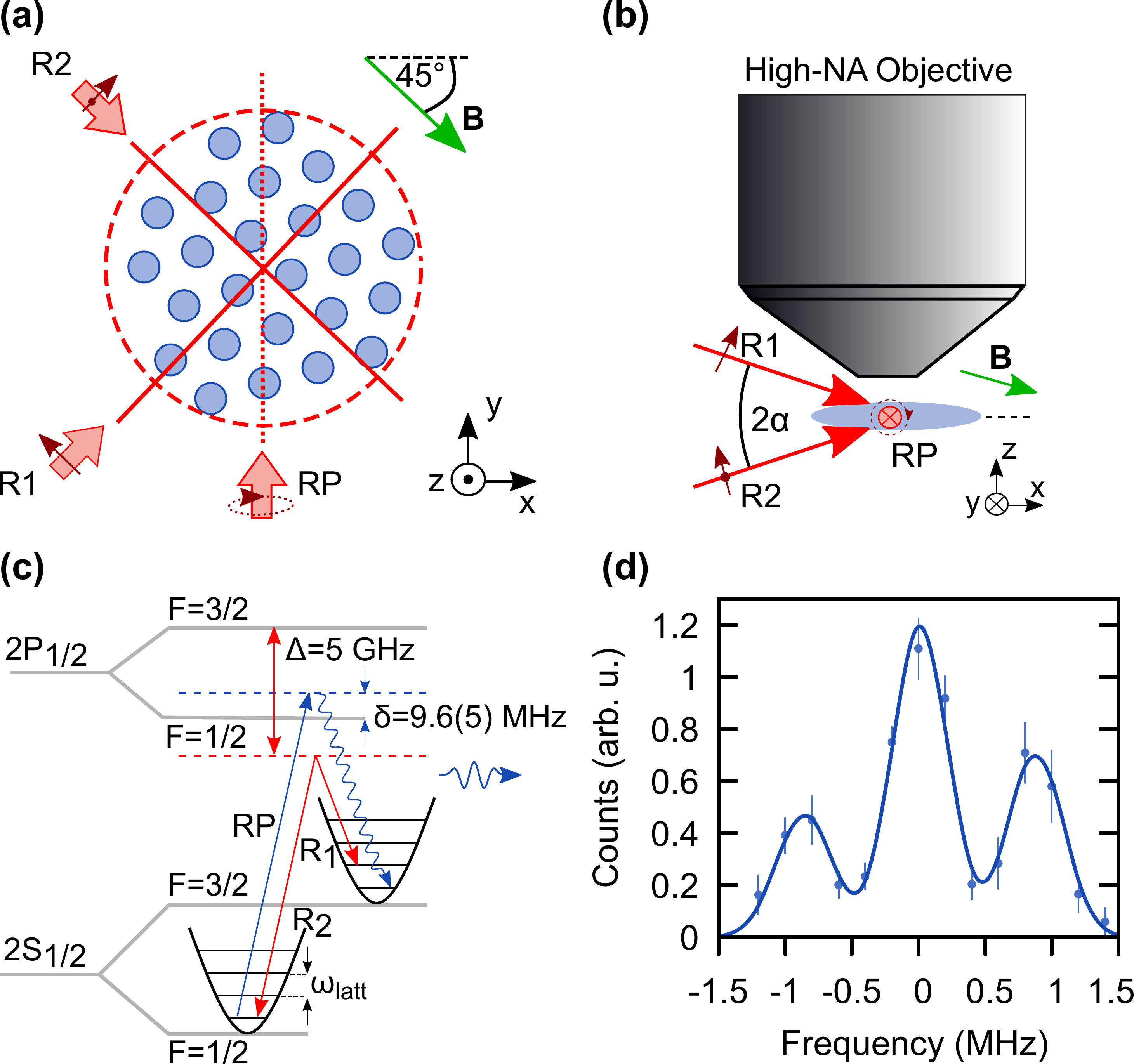}
	\caption{ 
{\bf Raman sideband cooling.} 
(a), (b) Raman sideband cooling beam configuration. Blue dots mark the triangular lattice sites. The Raman repump beam propagates in the lattice plane ($x$-$y$ plane). The first Raman beam (R1) has horizontal linear polarization and propagates in negative $z$ direction with a shallow angle of $\alpha=7.5(2)^\circ$ relative to the lattice plane. The second Raman beam (R2) is perpendicular to R1 and consists of a mix of horizontally and vertically linear polarizations in a ratio of 4\,:\,1. Two green arrows show the projection of the magnetic field on $x$-$y$ plane and $x$-$z$ plane with angles of $-45^\circ$ and $-70^\circ$ relative to $x$ axis, respectively. (c) Raman sideband cooling transition scheme showing the levels connected by the Raman repump RP and the Raman beams R1 and R2 and the respective detunings $\delta$ and $\Delta$. 
(d) Raman sideband spectrum in the triangular lattice. The center peak is the carrier corresponding to hyperfine splitting in the ground state while the sidebands show the lattice vibrational spacing of $\omega_\text{latt}=2\pi\times\SI{870(20)}{\kilo\Hz}$. The amplitude ratio of the sidebands indicates an average number of vibrational quanta of $2^{+3}_{-1}$ in $x$ and $y$ direction. The dots represent experimental data and the solid line is a Gaussian fit. Error bars are the standard deviation of four repetitions.}
	\label{fig: Raman beams configuration and Raman cooling process}
\end{figure}

Further specifics of our Raman sideband cooling setup are described as follows. A two-photon Raman transition via the $D_1$ line transfers the atoms from $\ket{\textrm{2S$_{1/2}$ F = 3/2}}$ manifold to $\ket{\textrm{2S$_{1/2}$ F = 1/2}}$ while lowering the vibrational state by one. The incoming Raman cooling beam (R1) is locked \SI{5}{GHz} red-detuned to the $D_1$ line and is linearly polarized. It has a power of \SI{2.2}{\milli\watt} and a beam waist of \SI{100}{\micro\meter} on the atoms. After passing through the chamber, we use a double-pass configuration of an acousto-optic modulator (AOM) to generate the second Raman beam (R2) with a detuning of $\SI{228.2}{\mega\Hz}+\omega_\text{latt}/(2\pi)$ and 70\% efficiency. 
We choose the angles between the two Raman beams and relative to the lattice to obtain sufficient coupling in-plane as well as in the $z$ direction [Fig.~\ref{fig: Raman beams configuration and Raman cooling process}{\bf(a,b)}].
To determine the $\omega_\text{latt}$ we take sideband spectra by applying a pulse of both Raman beams directly after loading into the lattice, transferring a fraction of the atoms from $\ket{\textrm{2S$_{1/2}$ F = 1/2}}$ to $\ket{\textrm{2S$_{1/2}$ F = 3/2}}$. These atoms are then detected in absorption imaging and the sidebands show the lattice vibrational spacing of $\omega_\text{latt} = 2\pi\times\SI{870(20)}{\kilo\Hz}$ [Fig.~\ref{fig: Raman beams configuration and Raman cooling process}{\bf(d)}].
The Raman repump beam (RP) has a power of \SI{0.15}{\milli\watt} and a beam waist of \SI{500}{\micro\meter} at the focus on the atoms. It is locked \SI{9.6(5)}{\mega\Hz} blue-detuned to the $\ket{\textrm{2S$_{1/2}$ F = 1/2}}$ to the $\ket{\textrm{2P$_{1/2}$ F = 1/2}}$ atomic transition and is circularly polarized. The atoms excited by the Raman repump beam more likely decay down into the $\ket{\textrm{2S$_{1/2}$ F = 3/2}}$ state rather than the $\ket{\textrm{2S$_{1/2}$ F = 1/2}}$ state with a branching ratio of 8\,:\,1. Their vibrational state in the lattice remains mostly unchanged due to the Lamb-Dicke factor $\eta\equiv\sqrt{\hbar k_R^2/(2m\omega_\text{latt})}= 0.29$ in our experiment, where $k_R$ is the wavevector of the Raman repump light. There may be higher vibrational states excited in the vertical light sheet direction. However, due to the depth of the light sheet and the absence of nearby wells the atoms could tunnel to, the impact of elevated temperatures in the $z$ dimension on imaging fidelity is low and we can rely on coupling to other dimensions for cooling, which is provided by the small angle $\alpha$ [Fig.~\ref{fig: Raman beams configuration and Raman cooling process}{\bf(b)}].

The spatial configuration of the Raman cooling beams and repump beam is shown in Fig.~\ref{fig: Sketch of the lattice setup and Raman sideband imaging setup} and Fig.~\ref{fig: Raman beams configuration and Raman cooling process}. To get the best imaging result, we optimize the offset magnetic fields, leading to a magnetic field of 1.1(2)\,G rather than zero field. The parameters for all offset magnetic fields are shown in Fig.~\ref{fig: Raman beams configuration and Raman cooling process}. The lifetime of atoms under continuous Raman cooling in the triangular lattice is \SI{44(2)}{\second}, possibly limited by background gas collisions.

\section{High-resolution imaging}

To achieve imaging of $^{6}$Li atoms in the triangular lattice with single-site resolved sensitivity, a high-resolution imaging system is used to collect the fluorescence during Raman sideband cooling. The imaging system consists of a custom objective (54-25-25@671nm, Navitar) with a focal length of \SI{25}{\milli\meter} and a numerical aperture (NA) of 0.5, and an achromatic doublet (AC508-750-B) with a focal length of \SI{750}{\milli\meter}, leading to a theoretical magnification of 30. The measured magnification is 33. Scattered photons are detected with an exposure time of \SI{500}{\milli\second} by a low-noise scientific CMOS camera (Andor Zyla 4.2 plus) with quantum efficiency of 77\% and pixel size $6.5\times6.5$ \SI{}{\micro\meter}$^2$. The total transmission of imaging optics and narrow-band filters is $\sim$ 80\%, leading to a total photon collection efficiency of $\sim$ 5.4\%. From our pictures we conclude that we detect about 1000 photons per atom, corresponding to a scattering rate for each atom of about \SI{34}{\kilo\Hz}, calculated by dividing the number of photons per atom by the total collection efficiency and the exposure time.

To verify that individual lattice sites are well-resolved, it is necessary to check the point spread function of the system. For this purpose, we take pictures of dilute systems with many well-separated atoms. The point spread function (PSF) is obtained by overlapping and averaging approximately 800 single atoms. After azimuthal averaging we fit to a Gaussian function [Fig.~\ref{fig: 3}{\bf(a)}]. This reveals a full width at half maximum (FWHM) of \SI{720(18)}{\nano\meter}, consistent with our expectation of \SI{711}{\nano\meter}. From the first minimum of an Airy fit, we extract the resolution according to the Rayleigh criterion as \SI{818(8)}{\nano\meter} (4.3 pixels). This is smaller than our triangular lattice spacing of \SI{1003}{\nano\meter} (5.1 pixels), and we therefore resolve individual atoms in the triangular lattice without post-processing.

\section{Image reconstruction and analysis}

\begin{figure}[ht]
	\centering
	\includegraphics[width=\linewidth]{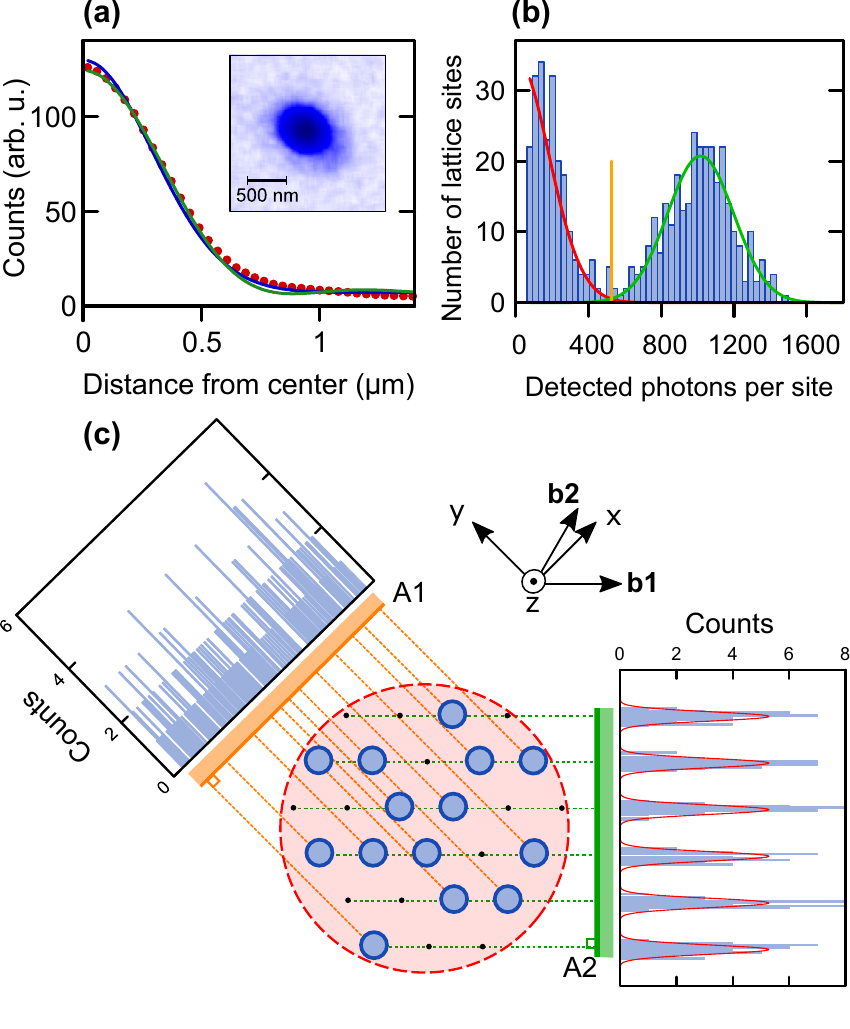}
	\caption{{\bf Image analysis.} (a) Point spread function. Azimuthal average of the point spread function ({red}), with Gaussian fit ({blue}), and Airy fit ({green}). The measured FWHM of \SI{720(18)}{\nano\meter} of the PSF is consistent with the FWHM of \SI{711}{\nano\meter} expected from the numerical aperture of the objective. The inset shows the PSF obtained by averaging isolated atoms. (b) Single atom count histogram. The left peak corresponds to empty sites and the right peak indicates sites occupied by single atoms. The threshold value between no atom and a single atom (vertical orange line) is determined as the intersection point of two Gaussian fits to background and atom signal distribution, respectively. The reconstruction error caused by the overlap is negligible compared to the observed hopping and loss. (c) Determining lattice angles and lattice constant. Single atom positions (exemplary shown in blue) are orthogonally projected onto a line of varying angle, here exemplified by A1 and A2. For each angle, a histogram of the projected positions is depicted. At the lattice angle, the experimental histogram has perfect contrast (right graph) corresponding to lattice vector $b1$. At other angles, for example A2, there is almost no structure in the histograms.}
	\label{fig: 3}
\end{figure}

We apply a reconstruction algorithm to extract a digitized occupation matrix of the lattice \cite{Sherson2010}. In order to obtain the geometric parameters of the triangular lattice, we determine lattice angles and lattice constants. This relies on identifying individual isolated atoms and determining their center via Gaussian fits. Then, we project the coordinates of isolated atoms onto an axis with varying rotational angle in the lattice plane. By introducing equidistant bins on this axis we generate a histogram of atom projections [Fig.~\ref{fig: 3}{\bf (c)}]. If the rotation angle is very close to the lattice angle, the histogram shows multiple peaks with minimal width and the separation between the peaks is related to the lattice constant. The angles with respect to $x$ axis for both lattice vectors are determined with high precision to $-45.85(3)^\circ$ and $13.51(1)^\circ$, leading to 59.36(3)$^\circ$ between the lattice vectors. These lattice angles allow us to extract the precision of the relative angles between the lattice beams to 120.0(6)$^{\circ}$. The lattice constants in pixels are $5.09(9)$ and $5.10(4)$, consistent with a symmetric triangular lattice.

While the lattice angles only vary because of alignment changes, the phase of the lattice usually drifts due to thermal effects. To estimate the phase in a picture, we generate the lattice structure and compare it to the position of isolated single atoms and then measure the phase difference between every single atom and the nearest lattice site. With knowledge of the lattice angle, lattice constant, and lattice phase, the exact position of all lattice sites in image coordinates is revealed.

To obtain the occupation of each lattice site, we simultaneously fit 2D Gaussian functions to all lattice sites with significant signal of more than about hundred detected photons per site. The resulting histogram of all Gaussian amplitudes in Fig.~\ref{fig: 3}{\bf (b)} shows a well-separated peak of single atom signal. Due to light-induced collisions, doubly occupied sites are detected as empty sites \cite{Anderson1994,Fuhrmanek2012,Endres2013}. From the histogram, we obtain an optimized threshold between the signal of no atom and single atoms to decide if a lattice site is occupied. As a result of the reconstruction, a matrix with entries zero (empty) or one (occupied) is generated. To handle the triangular lattice structure, we interpret it as a square lattice with diagonal tunneling, sheared by 30$^\circ$.

\subsection{Imaging fidelity}

\begin{figure}[ht]
	\includegraphics[width=\linewidth]{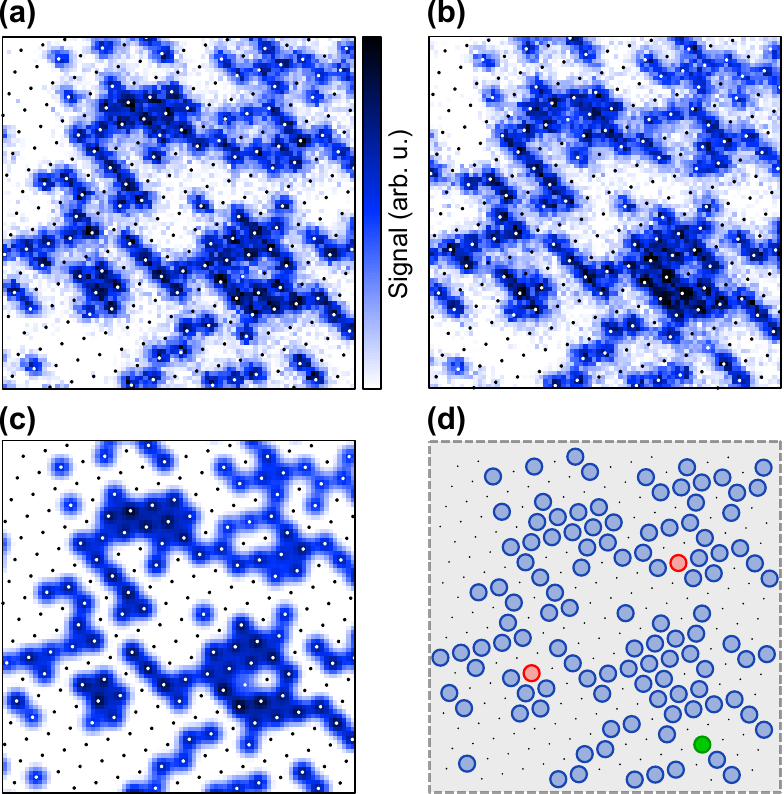}
	\caption{{\bf Imaging fidelity.} (a), (b) Two adjacent images of individual $^{6}$Li atoms ({white} dots) in a triangular lattice ({black} dots) imaged with \SI{500}{\milli\second} exposure and separation of \SI{50}{\milli\second}. (c) Reconstructed occupation of picture (a) convolved with the PSF. (d) Hopping and loss during imaging, stationary atoms ({blue}), hopped atoms ({green}) and lost atoms ({red}).}
	\label{fig: 4}
\end{figure}

We evaluate the imaging fidelity by taking a series of five images with \SI{500}{\milli\second} exposure time each and \SI{50}{\milli\second} separation in between. Hopping and loss rate are estimated by comparing two adjacent images [Fig.~\ref{fig: 4}]. The hopping rate is defined by the fraction of sites detected as occupied in the second image only while the loss rate is given by the fraction of atoms lost from picture to picture. We define the imaging fidelity by the fraction of atoms that remained in the same lattice sites. Our single-site imaging has a field of view of $90\times\SI{90}{\micro\meter^2}$, with good hopping rates in a region of $25\times\SI{25}{\micro\meter^2}$ in the center of the atom distribution, which includes 625 lattice sites. 
We obtain a hopping rate of 2.0(2)\% and a loss rate of 0.4(2)\% at a detected occupation of up to 50\% by averaging over twelve pairs of pictures. This demonstrates an imaging fidelity of 97.6(3)\%. The detected density is reduced by light-induced pair-wise losses at the beginning of the first image which lowers the density of the loaded Fermi gas from an initial density of approximately 1.3 atoms per lattice site, determined independently by high-field absorption imaging.

\subsection{Comparison to square lattices}

\begin{figure}
	\centering
	\includegraphics[width=\linewidth]{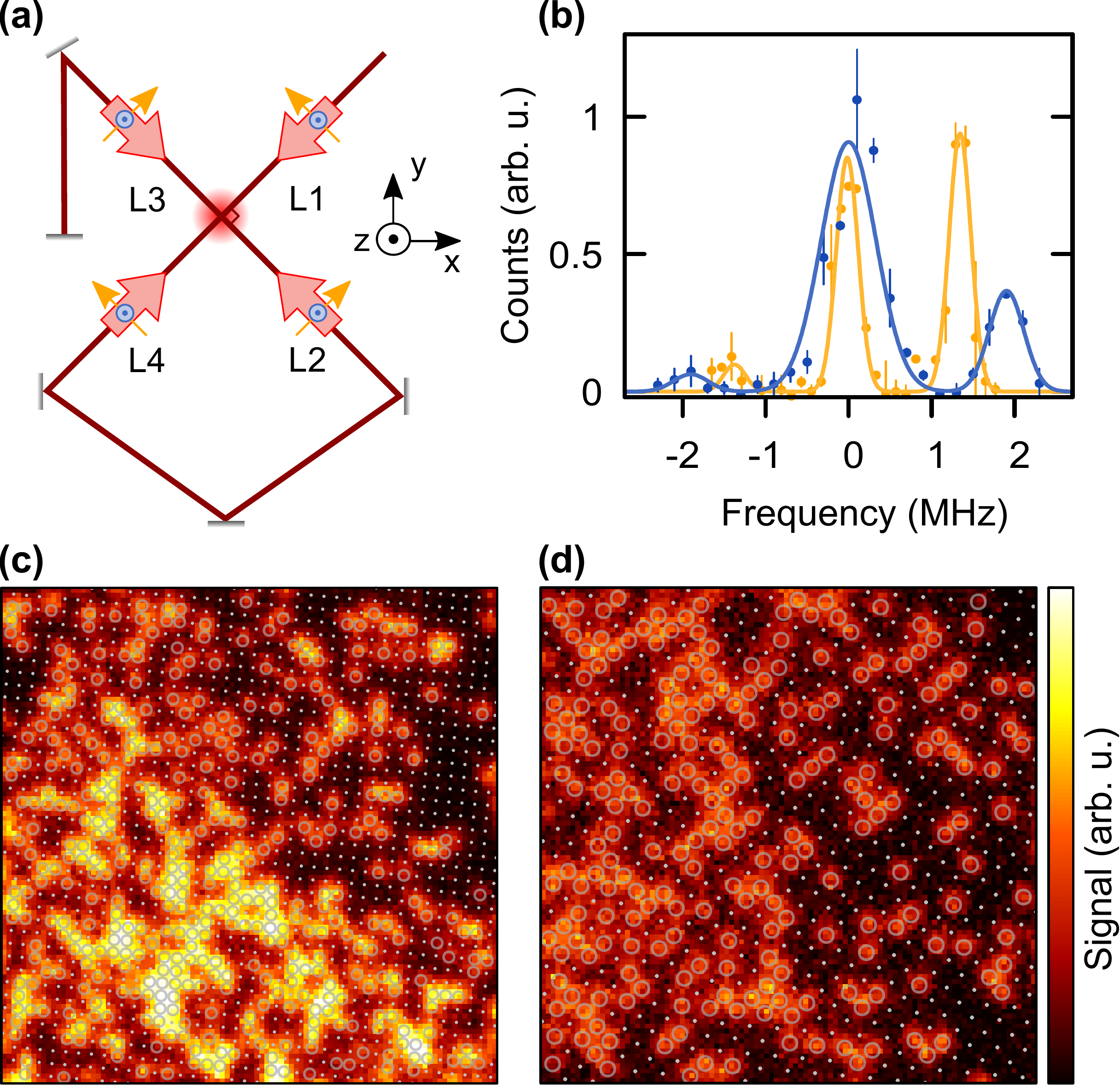}
	\caption{{\bf Comparison to square lattices.}  (a) Square lattice setup. Orange and blue arrows denote polarizations of \SI{532}{\nano\meter} and \SI{752}{\nano\meter} spacing square lattices, respectively. (b) Raman sideband spectra in \SI{532}{\nano\meter} spacing (orange) and \SI{752}{\nano\meter} spacing (blue) square lattice. The dots denote experimental data and solid lines are Gaussian fits. The sidebands are at \SI{1.36(2)}{\mega\Hz} and \SI{1.90(4)}{\mega\Hz} for \SI{532}{\nano\meter} and \SI{752}{\nano\meter} square lattices, respectively. The asymmetry of the sidebands shows that the atoms are predominantly in the 2d vibrational ground state after loading into the lattice. We find an average number of vibrational quanta per dimension in 2d of $0.1(1)$ in the \SI{532}{\nano\meter} lattice and $0.2^{+0.8}_{-0.2}$ in the \SI{752}{\nano\meter} lattice. (c), (d) Single-site-resolved images of $^{6}$Li atoms with lattice structure overlay in the \SI{532}{\nano\meter} spacing and \SI{752}{\nano\meter} spacing lattice, respectively. The gray circles indicate occupied lattice sites. For the \SI{532}{\nano\meter} lattice, the Raman configuration is the same as for the triangular, but for the \SI{752}{\nano\meter} lattice we use counter-propagating Raman beams. }
	\label{fig: Single atoms and Raman sideband in square lattice}
\end{figure}

In addition to the triangular lattice, we implemented a versatile square lattice at the same experimental setup which can be superimposed with the triangular lattice. The square lattice setup can be used at \SI{532}{\nano\meter} or \SI{752}{\nano\meter} lattice spacing. We create the square lattices using the recycled lattice setup as described in refs.~\cite{Sebby2006, Brown2017} [Fig.~\ref{fig: Single atoms and Raman sideband in square lattice}{\bf(a)}]. For vertical polarization, four-beam interference leads to a \SI{752}{\nano\meter} spacing lattice, while an in-plane polarization creates a \SI{532}{\nano\meter} spacing lattice. The power of the four passes is \SI{41}{\watt}, \SI{39}{\watt}, \SI{37}{\watt} and \SI{36}{\watt}, respectively, with a Gaussian beam waist of \SI{70}{\micro\meter}. The trap depths are 1900\,$E_r^{\text{532nm}}$ and 7500\,$E_r^{\text{752nm}}$ and trap frequencies are \SI{1.36(2)}{\mega\Hz} and \SI{1.90(4)}{\mega\Hz} for the \SI{532}{\nano\meter} and \SI{752}{\nano\meter} spacing lattices, respectively [Fig.~\ref{fig: Single atoms and Raman sideband in square lattice}{\bf(b)}]. 

The square lattices have smaller lattice spacing than the triangular lattice, however, our reconstruction algorithm is able to determine the lattice occupation with an error only limited by the observed hopping and loss [Fig.~\ref{fig: Single atoms and Raman sideband in square lattice}{\bf(c,d)}]. We confirmed this by comparing different fitting subroutines which lead to differences much smaller than the imaging infidelity. The \SI{532}{\nano\meter} spacing lattice is imaged using the same Raman cooling configuration as the triangular lattice, while for the \SI{752}{\nano\meter} square lattice the Raman beam R2 is the retroreflection of the incoming Raman beam R1, instead of the orthogonal configuration described above. For the triangular and \SI{532}{\nano\meter} spacing square lattices with smaller trap frequencies, we observed that the orthogonal Raman beam configuration is necessary, but for trap frequencies beyond \SI{1.5}{\mega\Hz}, the retroreflected configuration works well. The square lattices have imaging fidelities of 84(3)\% and 97(1)\%, with detected filling up to 50\%, in \SI{532}{\nano\meter} and \SI{752}{\nano\meter} spacing lattices, respectively. 

Our imaging fidelity in the \SI{532}{\nano\meter} spacing lattice is slightly lower than observed previously in a three-dimensional \SI{532}{\nano\meter} spacing lattice, possibly caused by our weaker $z$ confinement \cite{Omran2015}. However, the imaging fidelity in the \SI{752}{\nano\meter} spacing lattice is comparable with previous results \cite{Brown2017}. Due to the large sideband frequency in our \SI{752}{\nano\meter} spacing lattice, it would be possible to double the system size while maintaining sufficient lattice depth for high-fidelity imaging.
Superimposing the triangular lattice with the square lattice can form a two-dimensional quasi-crystalline lattice \cite{Sbroscia2020}, which could be used to study many-body localization in a non-separable two-dimensional quasi-periodic lattice. Our setup is ready to superimpose both lattices by splitting the laser power between both simultaneously realized optical paths and will be capable to study such systems on a single-atom level.

\section{Conclusion and Outlook}

We have presented the first single-site imaging of ultracold fermionic atoms in a triangular lattice, demonstrating a state-of-the-art imaging fidelity of 97.6(3)\%. Our triangular lattice with spacing \SI{1003}{\nano\meter} enables fast tunneling rates of $\sim$\,\SI{700}{\Hz} in the strongly interacting Hubbard-regime. The interactions are tunable via the Feshbach resonance in lithium and are only limited by multi-band effects.
In our current configuration, we estimate that about 20 sites are in the Hubbard regime when loading the lattice at maximum light sheet depth, making it very challenging to observe interaction effects.
Through the addition of a vertical lattice we will increase the vertical confinement to suppress multi-band effects to obtain Hubbard systems of several hundred atoms.
Our platform will enable studies of the Fermi-Hubbard model in the triangular lattice and, in the limit of strong interactions, the triangular Heisenberg spin model. By varying the polarizations of the three lattice beams, we can adiabatically change the triangular lattice between symmetric and asymmetric tunneling configurations, enabling the study of the complete tunneling-imbalance parameter space.
Furthermore, the platform is ideally suited to directly measure emergent quantum correlations, study signatures of frustration and possibly even detect signatures of quantum spin liquids, depending on the lowest entropy states that can be prepared.
It will become possible to study spin-spin correlations in analogy with results for square lattices in the Mott-insulating regime \cite{Parsons2016,Boll2016,Cheuk2016,Brown2017}, possibly detecting the cross-over from three-sublattice order to a non-magnetic state \cite{Shirakawa2017,Szasz2020}.
Even at temperatures previously reached in ultracold Hubbard simulations, remnants of chiral correlations could be detected which would directly show time-reversal symmetry breaking \cite{Wen1989,Shirakawa2017}. Our new quantum gas microscope platform provides the basis for measuring these three-point correlations. Moreover, the triangular lattice Hubbard model exhibits kinetic frustration, which could be probed using a transient grating approach \cite{Brown2019,Vranic2020} or by detecting magnon-hole bound states. The bound states in the triangular lattice have binding energies that scale with the tunneling energy and are therefore at experimentally accessible temperatures \cite{Zhang2018}.

\raggedbottom
\begin{acknowledgments}
This work was supported by the University of Virginia.
We thank W. S. Bakr, S. S. Kondov and C. A. Sackett for comments on the manuscript and acknowledge discussions with D. Mitra, P. T. Brown, and E. Guardado-Sanchez. We thank J. W. Kim for early contributions to the experiment and S. Kuhr for sharing the initial code base for experiment control and reconstruction software which we extended for generalized lattice geometries.
\end{acknowledgments}





\begin{thebibliography}{47}%
\makeatletter
\providecommand \@ifxundefined [1]{%
 \@ifx{#1\undefined}
}%
\providecommand \@ifnum [1]{%
 \ifnum #1\expandafter \@firstoftwo
 \else \expandafter \@secondoftwo
 \fi
}%
\providecommand \@ifx [1]{%
 \ifx #1\expandafter \@firstoftwo
 \else \expandafter \@secondoftwo
 \fi
}%
\providecommand \natexlab [1]{#1}%
\providecommand \enquote  [1]{``#1''}%
\providecommand \bibnamefont  [1]{#1}%
\providecommand \bibfnamefont [1]{#1}%
\providecommand \citenamefont [1]{#1}%
\providecommand \href@noop [0]{\@secondoftwo}%
\providecommand \href [0]{\begingroup \@sanitize@url \@href}%
\providecommand \@href[1]{\@@startlink{#1}\@@href}%
\providecommand \@@href[1]{\endgroup#1\@@endlink}%
\providecommand \@sanitize@url [0]{\catcode `\\12\catcode `\$12\catcode
  `\&12\catcode `\#12\catcode `\^12\catcode `\_12\catcode `\%12\relax}%
\providecommand \@@startlink[1]{}%
\providecommand \@@endlink[0]{}%
\providecommand \url  [0]{\begingroup\@sanitize@url \@url }%
\providecommand \@url [1]{\endgroup\@href {#1}{\urlprefix }}%
\providecommand \urlprefix  [0]{URL }%
\providecommand \Eprint [0]{\href }%
\providecommand \doibase [0]{https://doi.org/}%
\providecommand \selectlanguage [0]{\@gobble}%
\providecommand \bibinfo  [0]{\@secondoftwo}%
\providecommand \bibfield  [0]{\@secondoftwo}%
\providecommand \translation [1]{[#1]}%
\providecommand \BibitemOpen [0]{}%
\providecommand \bibitemStop [0]{}%
\providecommand \bibitemNoStop [0]{.\EOS\space}%
\providecommand \EOS [0]{\spacefactor3000\relax}%
\providecommand \BibitemShut  [1]{\csname bibitem#1\endcsname}%
\let\auto@bib@innerbib\@empty
\bibitem [{\citenamefont {Wannier}(1950)}]{Wannier1950}%
  \BibitemOpen
  \bibfield  {author} {\bibinfo {author} {\bibfnamefont {G.~H.}\ \bibnamefont
  {Wannier}},\ }\bibfield  {title} {\bibinfo {title} {Antiferromagnetism. {The}
  {Triangular} {Ising} {Net}},\ }\href {https://doi.org/10.1103/PhysRev.79.357}
  {\bibfield  {journal} {\bibinfo  {journal} {Phys. Rev.}\ }\textbf {\bibinfo
  {volume} {79}},\ \bibinfo {pages} {357} (\bibinfo {year} {1950})}\BibitemShut
  {NoStop}%
\bibitem [{\citenamefont {Anderson}(1987)}]{Anderson1987}%
  \BibitemOpen
  \bibfield  {author} {\bibinfo {author} {\bibfnamefont {P.~W.}\ \bibnamefont
  {Anderson}},\ }\bibfield  {title} {\bibinfo {title} {The {Resonating}
  {Valence} {Bond} {State} in {La$_2$CuO$_4$} and {Superconductivity}},\ }\href
  {https://doi.org/10.1126/science.235.4793.1196} {\bibfield  {journal}
  {\bibinfo  {journal} {Science}\ }\textbf {\bibinfo {volume} {235}},\ \bibinfo
  {pages} {1196} (\bibinfo {year} {1987})}\BibitemShut {NoStop}%
\bibitem [{\citenamefont {Wen}\ \emph {et~al.}(1989)\citenamefont {Wen},
  \citenamefont {Wilczek},\ and\ \citenamefont {Zee}}]{Wen1989}%
  \BibitemOpen
  \bibfield  {author} {\bibinfo {author} {\bibfnamefont {X.~G.}\ \bibnamefont
  {Wen}}, \bibinfo {author} {\bibfnamefont {F.}~\bibnamefont {Wilczek}},\ and\
  \bibinfo {author} {\bibfnamefont {A.}~\bibnamefont {Zee}},\ }\bibfield
  {title} {\bibinfo {title} {Chiral spin states and superconductivity},\ }\href
  {https://doi.org/10.1103/PhysRevB.39.11413} {\bibfield  {journal} {\bibinfo
  {journal} {Phys. Rev. B}\ }\textbf {\bibinfo {volume} {39}},\ \bibinfo
  {pages} {11413} (\bibinfo {year} {1989})}\BibitemShut {NoStop}%
\bibitem [{\citenamefont {Balents}(2010)}]{Balents2010}%
  \BibitemOpen
  \bibfield  {author} {\bibinfo {author} {\bibfnamefont {L.}~\bibnamefont
  {Balents}},\ }\bibfield  {title} {\bibinfo {title} {Spin liquids in
  frustrated magnets},\ }\href {https://doi.org/10.1038/nature08917} {\bibfield
   {journal} {\bibinfo  {journal} {Nature}\ }\textbf {\bibinfo {volume}
  {464}},\ \bibinfo {pages} {199} (\bibinfo {year} {2010})}\BibitemShut
  {NoStop}%
\bibitem [{\citenamefont {Batista}\ \emph {et~al.}(2016)\citenamefont
  {Batista}, \citenamefont {Lin}, \citenamefont {Hayami},\ and\ \citenamefont
  {Kamiya}}]{Batista2016}%
  \BibitemOpen
  \bibfield  {author} {\bibinfo {author} {\bibfnamefont {C.~D.}\ \bibnamefont
  {Batista}}, \bibinfo {author} {\bibfnamefont {S.-Z.}\ \bibnamefont {Lin}},
  \bibinfo {author} {\bibfnamefont {S.}~\bibnamefont {Hayami}},\ and\ \bibinfo
  {author} {\bibfnamefont {Y.}~\bibnamefont {Kamiya}},\ }\bibfield  {title}
  {\bibinfo {title} {Frustration and chiral orderings in correlated electron
  systems},\ }\href {https://doi.org/10.1088/0034-4885/79/8/084504} {\bibfield
  {journal} {\bibinfo  {journal} {Rep. Prog. Phys.}\ }\textbf {\bibinfo
  {volume} {79}},\ \bibinfo {pages} {084504} (\bibinfo {year}
  {2016})}\BibitemShut {NoStop}%
\bibitem [{\citenamefont {Zhou}\ \emph {et~al.}(2017)\citenamefont {Zhou},
  \citenamefont {Kanoda},\ and\ \citenamefont {Ng}}]{Zhou2017}%
  \BibitemOpen
  \bibfield  {author} {\bibinfo {author} {\bibfnamefont {Y.}~\bibnamefont
  {Zhou}}, \bibinfo {author} {\bibfnamefont {K.}~\bibnamefont {Kanoda}},\ and\
  \bibinfo {author} {\bibfnamefont {T.-K.}\ \bibnamefont {Ng}},\ }\bibfield
  {title} {\bibinfo {title} {Quantum spin liquid states},\ }\href
  {https://doi.org/10.1103/RevModPhys.89.025003} {\bibfield  {journal}
  {\bibinfo  {journal} {Rev. Mod. Phys.}\ }\textbf {\bibinfo {volume} {89}},\
  \bibinfo {pages} {025003} (\bibinfo {year} {2017})}\BibitemShut {NoStop}%
\bibitem [{\citenamefont {Yoshioka}\ \emph {et~al.}(2009)\citenamefont
  {Yoshioka}, \citenamefont {Koga},\ and\ \citenamefont
  {Kawakami}}]{Yoshioka2009}%
  \BibitemOpen
  \bibfield  {author} {\bibinfo {author} {\bibfnamefont {T.}~\bibnamefont
  {Yoshioka}}, \bibinfo {author} {\bibfnamefont {A.}~\bibnamefont {Koga}},\
  and\ \bibinfo {author} {\bibfnamefont {N.}~\bibnamefont {Kawakami}},\
  }\bibfield  {title} {\bibinfo {title} {Quantum phase transitions in the
  {Hubbard} model on a triangular lattice},\ }\href
  {https://doi.org/10.1103/PhysRevLett.103.036401} {\bibfield  {journal}
  {\bibinfo  {journal} {Phys. Rev. Lett.}\ }\textbf {\bibinfo {volume} {103}},\
  \bibinfo {pages} {036401} (\bibinfo {year} {2009})}\BibitemShut {NoStop}%
\bibitem [{\citenamefont {Shirakawa}\ \emph {et~al.}(2017)\citenamefont
  {Shirakawa}, \citenamefont {Tohyama}, \citenamefont {Kokalj}, \citenamefont
  {Sota},\ and\ \citenamefont {Yunoki}}]{Shirakawa2017}%
  \BibitemOpen
  \bibfield  {author} {\bibinfo {author} {\bibfnamefont {T.}~\bibnamefont
  {Shirakawa}}, \bibinfo {author} {\bibfnamefont {T.}~\bibnamefont {Tohyama}},
  \bibinfo {author} {\bibfnamefont {J.}~\bibnamefont {Kokalj}}, \bibinfo
  {author} {\bibfnamefont {S.}~\bibnamefont {Sota}},\ and\ \bibinfo {author}
  {\bibfnamefont {S.}~\bibnamefont {Yunoki}},\ }\bibfield  {title} {\bibinfo
  {title} {Ground state phase diagram of the triangular lattice {Hubbard} model
  by density matrix renormalization group method},\ }\href
  {https://doi.org/10.1103/PhysRevB.96.205130} {\bibfield  {journal} {\bibinfo
  {journal} {Phys. Rev. B}\ }\textbf {\bibinfo {volume} {96}},\ \bibinfo
  {pages} {205130} (\bibinfo {year} {2017})}\BibitemShut {NoStop}%
\bibitem [{\citenamefont {Szasz}\ \emph {et~al.}(2020)\citenamefont {Szasz},
  \citenamefont {Motruk}, \citenamefont {Zaletel},\ and\ \citenamefont
  {Moore}}]{Szasz2020}%
  \BibitemOpen
  \bibfield  {author} {\bibinfo {author} {\bibfnamefont {A.}~\bibnamefont
  {Szasz}}, \bibinfo {author} {\bibfnamefont {J.}~\bibnamefont {Motruk}},
  \bibinfo {author} {\bibfnamefont {M.~P.}\ \bibnamefont {Zaletel}},\ and\
  \bibinfo {author} {\bibfnamefont {J.~E.}\ \bibnamefont {Moore}},\ }\bibfield
  {title} {\bibinfo {title} {Chiral spin liquid phase of the triangular lattice
  {Hubbard} model: {A} density matrix renormalization group study},\ }\href
  {https://doi.org/10.1103/PhysRevX.10.021042} {\bibfield  {journal} {\bibinfo
  {journal} {Phys. Rev. X}\ }\textbf {\bibinfo {volume} {10}},\ \bibinfo
  {pages} {021042} (\bibinfo {year} {2020})}\BibitemShut {NoStop}%
\bibitem [{\citenamefont {Greif}\ \emph {et~al.}(2013)\citenamefont {Greif},
  \citenamefont {Uehlinger}, \citenamefont {Jotzu}, \citenamefont {Tarruell},\
  and\ \citenamefont {Esslinger}}]{Greif2013}%
  \BibitemOpen
  \bibfield  {author} {\bibinfo {author} {\bibfnamefont {D.}~\bibnamefont
  {Greif}}, \bibinfo {author} {\bibfnamefont {T.}~\bibnamefont {Uehlinger}},
  \bibinfo {author} {\bibfnamefont {G.}~\bibnamefont {Jotzu}}, \bibinfo
  {author} {\bibfnamefont {L.}~\bibnamefont {Tarruell}},\ and\ \bibinfo
  {author} {\bibfnamefont {T.}~\bibnamefont {Esslinger}},\ }\bibfield  {title}
  {\bibinfo {title} {Short-range quantum magnetism of ultracold fermions in an
  optical lattice},\ }\href {https://doi.org/10.1126/science.1236362}
  {\bibfield  {journal} {\bibinfo  {journal} {Science}\ }\textbf {\bibinfo
  {volume} {340}},\ \bibinfo {pages} {1307} (\bibinfo {year}
  {2013})}\BibitemShut {NoStop}%
\bibitem [{\citenamefont {Hart}\ \emph {et~al.}(2015)\citenamefont {Hart},
  \citenamefont {Duarte}, \citenamefont {Yang}, \citenamefont {Liu},
  \citenamefont {Paiva}, \citenamefont {Khatami}, \citenamefont {Scalettar},
  \citenamefont {Trivedi}, \citenamefont {Huse},\ and\ \citenamefont
  {Hulet}}]{Hart2015}%
  \BibitemOpen
  \bibfield  {author} {\bibinfo {author} {\bibfnamefont {R.~A.}\ \bibnamefont
  {Hart}}, \bibinfo {author} {\bibfnamefont {P.~M.}\ \bibnamefont {Duarte}},
  \bibinfo {author} {\bibfnamefont {T.-L.}\ \bibnamefont {Yang}}, \bibinfo
  {author} {\bibfnamefont {X.}~\bibnamefont {Liu}}, \bibinfo {author}
  {\bibfnamefont {T.}~\bibnamefont {Paiva}}, \bibinfo {author} {\bibfnamefont
  {E.}~\bibnamefont {Khatami}}, \bibinfo {author} {\bibfnamefont {R.~T.}\
  \bibnamefont {Scalettar}}, \bibinfo {author} {\bibfnamefont {N.}~\bibnamefont
  {Trivedi}}, \bibinfo {author} {\bibfnamefont {D.~A.}\ \bibnamefont {Huse}},\
  and\ \bibinfo {author} {\bibfnamefont {R.~G.}\ \bibnamefont {Hulet}},\
  }\bibfield  {title} {\bibinfo {title} {Observation of antiferromagnetic
  correlations in the {Hubbard} model with ultracold atoms},\ }\href
  {https://doi.org/10.1038/nature14223} {\bibfield  {journal} {\bibinfo
  {journal} {Nature}\ }\textbf {\bibinfo {volume} {519}},\ \bibinfo {pages}
  {211} (\bibinfo {year} {2015})}\BibitemShut {NoStop}%
\bibitem [{\citenamefont {Drewes}\ \emph {et~al.}(2017)\citenamefont {Drewes},
  \citenamefont {Miller}, \citenamefont {Cocchi}, \citenamefont {Chan},
  \citenamefont {Wurz}, \citenamefont {Gall}, \citenamefont {Pertot},
  \citenamefont {Brennecke},\ and\ \citenamefont {K\"ohl}}]{Drewes2016b}%
  \BibitemOpen
  \bibfield  {author} {\bibinfo {author} {\bibfnamefont {J.~H.}\ \bibnamefont
  {Drewes}}, \bibinfo {author} {\bibfnamefont {L.~A.}\ \bibnamefont {Miller}},
  \bibinfo {author} {\bibfnamefont {E.}~\bibnamefont {Cocchi}}, \bibinfo
  {author} {\bibfnamefont {C.~F.}\ \bibnamefont {Chan}}, \bibinfo {author}
  {\bibfnamefont {N.}~\bibnamefont {Wurz}}, \bibinfo {author} {\bibfnamefont
  {M.}~\bibnamefont {Gall}}, \bibinfo {author} {\bibfnamefont {D.}~\bibnamefont
  {Pertot}}, \bibinfo {author} {\bibfnamefont {F.}~\bibnamefont {Brennecke}},\
  and\ \bibinfo {author} {\bibfnamefont {M.}~\bibnamefont {K\"ohl}},\
  }\bibfield  {title} {\bibinfo {title} {Antiferromagnetic correlations in
  two-dimensional fermionic {Mott}-insulating and metallic phases},\ }\href
  {https://doi.org/10.1103/PhysRevLett.118.170401} {\bibfield  {journal}
  {\bibinfo  {journal} {Phys. Rev. Lett.}\ }\textbf {\bibinfo {volume} {118}},\
  \bibinfo {pages} {170401} (\bibinfo {year} {2017})}\BibitemShut {NoStop}%
\bibitem [{\citenamefont {Parsons}\ \emph {et~al.}(2016)\citenamefont
  {Parsons}, \citenamefont {Mazurenko}, \citenamefont {Chiu}, \citenamefont
  {Ji}, \citenamefont {Greif},\ and\ \citenamefont {Greiner}}]{Parsons2016}%
  \BibitemOpen
  \bibfield  {author} {\bibinfo {author} {\bibfnamefont {M.~F.}\ \bibnamefont
  {Parsons}}, \bibinfo {author} {\bibfnamefont {A.}~\bibnamefont {Mazurenko}},
  \bibinfo {author} {\bibfnamefont {C.~S.}\ \bibnamefont {Chiu}}, \bibinfo
  {author} {\bibfnamefont {G.}~\bibnamefont {Ji}}, \bibinfo {author}
  {\bibfnamefont {D.}~\bibnamefont {Greif}},\ and\ \bibinfo {author}
  {\bibfnamefont {M.}~\bibnamefont {Greiner}},\ }\bibfield  {title} {\bibinfo
  {title} {Site-resolved measurement of the spin-correlation function in the
  {Fermi}-{Hubbard} model},\ }\href {https://doi.org/10.1126/science.aag1430}
  {\bibfield  {journal} {\bibinfo  {journal} {Science}\ }\textbf {\bibinfo
  {volume} {353}},\ \bibinfo {pages} {1253} (\bibinfo {year}
  {2016})}\BibitemShut {NoStop}%
\bibitem [{\citenamefont {Boll}\ \emph {et~al.}(2016)\citenamefont {Boll},
  \citenamefont {Hilker}, \citenamefont {Salomon}, \citenamefont {Omran},
  \citenamefont {Nespolo}, \citenamefont {Pollet}, \citenamefont {Bloch},\ and\
  \citenamefont {Gross}}]{Boll2016}%
  \BibitemOpen
  \bibfield  {author} {\bibinfo {author} {\bibfnamefont {M.}~\bibnamefont
  {Boll}}, \bibinfo {author} {\bibfnamefont {T.~A.}\ \bibnamefont {Hilker}},
  \bibinfo {author} {\bibfnamefont {G.}~\bibnamefont {Salomon}}, \bibinfo
  {author} {\bibfnamefont {A.}~\bibnamefont {Omran}}, \bibinfo {author}
  {\bibfnamefont {J.}~\bibnamefont {Nespolo}}, \bibinfo {author} {\bibfnamefont
  {L.}~\bibnamefont {Pollet}}, \bibinfo {author} {\bibfnamefont
  {I.}~\bibnamefont {Bloch}},\ and\ \bibinfo {author} {\bibfnamefont
  {C.}~\bibnamefont {Gross}},\ }\bibfield  {title} {\bibinfo {title} {Spin- and
  density-resolved microscopy of antiferromagnetic correlations in
  {Fermi-Hubbard} chains},\ }\href {https://doi.org/10.1126/science.aag1635}
  {\bibfield  {journal} {\bibinfo  {journal} {Science}\ }\textbf {\bibinfo
  {volume} {353}},\ \bibinfo {pages} {1257} (\bibinfo {year}
  {2016})}\BibitemShut {NoStop}%
\bibitem [{\citenamefont {Cheuk}\ \emph {et~al.}(2016)\citenamefont {Cheuk},
  \citenamefont {Nichols}, \citenamefont {Lawrence}, \citenamefont {Okan},
  \citenamefont {Zhang}, \citenamefont {Khatami}, \citenamefont {Trivedi},
  \citenamefont {Paiva}, \citenamefont {Rigol},\ and\ \citenamefont
  {Zwierlein}}]{Cheuk2016}%
  \BibitemOpen
  \bibfield  {author} {\bibinfo {author} {\bibfnamefont {L.~W.}\ \bibnamefont
  {Cheuk}}, \bibinfo {author} {\bibfnamefont {M.~A.}\ \bibnamefont {Nichols}},
  \bibinfo {author} {\bibfnamefont {K.~R.}\ \bibnamefont {Lawrence}}, \bibinfo
  {author} {\bibfnamefont {M.}~\bibnamefont {Okan}}, \bibinfo {author}
  {\bibfnamefont {H.}~\bibnamefont {Zhang}}, \bibinfo {author} {\bibfnamefont
  {E.}~\bibnamefont {Khatami}}, \bibinfo {author} {\bibfnamefont
  {N.}~\bibnamefont {Trivedi}}, \bibinfo {author} {\bibfnamefont
  {T.}~\bibnamefont {Paiva}}, \bibinfo {author} {\bibfnamefont
  {M.}~\bibnamefont {Rigol}},\ and\ \bibinfo {author} {\bibfnamefont {M.~W.}\
  \bibnamefont {Zwierlein}},\ }\bibfield  {title} {\bibinfo {title}
  {Observation of spatial charge and spin correlations in the {2D}
  {Fermi-Hubbard} model},\ }\href {https://doi.org/10.1126/science.aag3349}
  {\bibfield  {journal} {\bibinfo  {journal} {Science}\ }\textbf {\bibinfo
  {volume} {353}},\ \bibinfo {pages} {1260} (\bibinfo {year}
  {2016})}\BibitemShut {NoStop}%
\bibitem [{\citenamefont {Brown}\ \emph {et~al.}(2017)\citenamefont {Brown},
  \citenamefont {Mitra}, \citenamefont {Guardado-Sanchez}, \citenamefont
  {Schauß}, \citenamefont {Kondov}, \citenamefont {Khatami}, \citenamefont
  {Paiva}, \citenamefont {Trivedi}, \citenamefont {Huse},\ and\ \citenamefont
  {Bakr}}]{Brown2017}%
  \BibitemOpen
  \bibfield  {author} {\bibinfo {author} {\bibfnamefont {P.~T.}\ \bibnamefont
  {Brown}}, \bibinfo {author} {\bibfnamefont {D.}~\bibnamefont {Mitra}},
  \bibinfo {author} {\bibfnamefont {E.}~\bibnamefont {Guardado-Sanchez}},
  \bibinfo {author} {\bibfnamefont {P.}~\bibnamefont {Schauß}}, \bibinfo
  {author} {\bibfnamefont {S.~S.}\ \bibnamefont {Kondov}}, \bibinfo {author}
  {\bibfnamefont {E.}~\bibnamefont {Khatami}}, \bibinfo {author} {\bibfnamefont
  {T.}~\bibnamefont {Paiva}}, \bibinfo {author} {\bibfnamefont
  {N.}~\bibnamefont {Trivedi}}, \bibinfo {author} {\bibfnamefont {D.~A.}\
  \bibnamefont {Huse}},\ and\ \bibinfo {author} {\bibfnamefont {W.~S.}\
  \bibnamefont {Bakr}},\ }\bibfield  {title} {\bibinfo {title} {Spin-imbalance
  in a {2D} {Fermi}-{Hubbard} system},\ }\href
  {https://doi.org/10.1126/science.aam7838} {\bibfield  {journal} {\bibinfo
  {journal} {Science}\ }\textbf {\bibinfo {volume} {357}},\ \bibinfo {pages}
  {1385} (\bibinfo {year} {2017})}\BibitemShut {NoStop}%
\bibitem [{\citenamefont {Gall}\ \emph {et~al.}(2021)\citenamefont {Gall},
  \citenamefont {Wurz}, \citenamefont {Samland}, \citenamefont {Chan},\ and\
  \citenamefont {K{\"o}hl}}]{Gall2021}%
  \BibitemOpen
  \bibfield  {author} {\bibinfo {author} {\bibfnamefont {M.}~\bibnamefont
  {Gall}}, \bibinfo {author} {\bibfnamefont {N.}~\bibnamefont {Wurz}}, \bibinfo
  {author} {\bibfnamefont {J.}~\bibnamefont {Samland}}, \bibinfo {author}
  {\bibfnamefont {C.~F.}\ \bibnamefont {Chan}},\ and\ \bibinfo {author}
  {\bibfnamefont {M.}~\bibnamefont {K{\"o}hl}},\ }\bibfield  {title} {\bibinfo
  {title} {Competing magnetic orders in a bilayer {Hubbard} model with
  ultracold atoms},\ }\href {https://doi.org/10.1038/s41586-020-03058-x}
  {\bibfield  {journal} {\bibinfo  {journal} {Nature}\ }\textbf {\bibinfo
  {volume} {589}},\ \bibinfo {pages} {40} (\bibinfo {year} {2021})}\BibitemShut
  {NoStop}%
\bibitem [{\citenamefont {Gross}\ and\ \citenamefont
  {Bloch}(2017)}]{Gross2017}%
  \BibitemOpen
  \bibfield  {author} {\bibinfo {author} {\bibfnamefont {C.}~\bibnamefont
  {Gross}}\ and\ \bibinfo {author} {\bibfnamefont {I.}~\bibnamefont {Bloch}},\
  }\bibfield  {title} {\bibinfo {title} {Quantum simulations with ultracold
  atoms in optical lattices},\ }\href {https://doi.org/10.1126/science.aal3837}
  {\bibfield  {journal} {\bibinfo  {journal} {Science}\ }\textbf {\bibinfo
  {volume} {357}},\ \bibinfo {pages} {995} (\bibinfo {year}
  {2017})}\BibitemShut {NoStop}%
\bibitem [{\citenamefont {Lewenstein}\ \emph {et~al.}(2007)\citenamefont
  {Lewenstein}, \citenamefont {Sanpera}, \citenamefont {Ahufinger},
  \citenamefont {Damski}, \citenamefont {Sen(De)},\ and\ \citenamefont
  {Sen}}]{Lewenstein2007}%
  \BibitemOpen
  \bibfield  {author} {\bibinfo {author} {\bibfnamefont {M.}~\bibnamefont
  {Lewenstein}}, \bibinfo {author} {\bibfnamefont {A.}~\bibnamefont {Sanpera}},
  \bibinfo {author} {\bibfnamefont {V.}~\bibnamefont {Ahufinger}}, \bibinfo
  {author} {\bibfnamefont {B.}~\bibnamefont {Damski}}, \bibinfo {author}
  {\bibfnamefont {A.}~\bibnamefont {Sen(De)}},\ and\ \bibinfo {author}
  {\bibfnamefont {U.}~\bibnamefont {Sen}},\ }\bibfield  {title} {\bibinfo
  {title} {Ultracold atomic gases in optical lattices: mimicking condensed
  matter physics and beyond},\ }\href
  {https://doi.org/10.1080/00018730701223200} {\bibfield  {journal} {\bibinfo
  {journal} {Adv. Phys.}\ }\textbf {\bibinfo {volume} {56}},\ \bibinfo {pages}
  {243} (\bibinfo {year} {2007})}\BibitemShut {NoStop}%
\bibitem [{\citenamefont {Bloch}\ \emph {et~al.}(2008)\citenamefont {Bloch},
  \citenamefont {Dalibard},\ and\ \citenamefont
  {Zwerger}}]{BlochDalibardZwerger2008}%
  \BibitemOpen
  \bibfield  {author} {\bibinfo {author} {\bibfnamefont {I.}~\bibnamefont
  {Bloch}}, \bibinfo {author} {\bibfnamefont {J.}~\bibnamefont {Dalibard}},\
  and\ \bibinfo {author} {\bibfnamefont {W.}~\bibnamefont {Zwerger}},\
  }\bibfield  {title} {\bibinfo {title} {{Many-body physics with ultracold
  gases}},\ }\href {https://doi.org/10.1103/RevModPhys.80.885} {\bibfield
  {journal} {\bibinfo  {journal} {Rev. Mod. Phys.}\ }\textbf {\bibinfo {volume}
  {80}},\ \bibinfo {pages} {885} (\bibinfo {year} {2008})}\BibitemShut
  {NoStop}%
\bibitem [{\citenamefont {Esslinger}(2010)}]{Esslinger2010}%
  \BibitemOpen
  \bibfield  {author} {\bibinfo {author} {\bibfnamefont {T.}~\bibnamefont
  {Esslinger}},\ }\bibfield  {title} {\bibinfo {title} {{Fermi-Hubbard} physics
  with atoms in an optical lattice},\ }\href
  {https://doi.org/10.1146/annurev-conmatphys-070909-104059} {\bibfield
  {journal} {\bibinfo  {journal} {Annu. Rev. Condens. Matter Phys.}\ }\textbf
  {\bibinfo {volume} {1}},\ \bibinfo {pages} {129} (\bibinfo {year}
  {2010})}\BibitemShut {NoStop}%
\bibitem [{\citenamefont {J\"ordens}\ \emph {et~al.}(2008)\citenamefont
  {J\"ordens}, \citenamefont {Strohmaier}, \citenamefont {G\"unter},
  \citenamefont {Moritz},\ and\ \citenamefont {Esslinger}}]{Joerdens2008}%
  \BibitemOpen
  \bibfield  {author} {\bibinfo {author} {\bibfnamefont {R.}~\bibnamefont
  {J\"ordens}}, \bibinfo {author} {\bibfnamefont {N.}~\bibnamefont
  {Strohmaier}}, \bibinfo {author} {\bibfnamefont {K.}~\bibnamefont
  {G\"unter}}, \bibinfo {author} {\bibfnamefont {H.}~\bibnamefont {Moritz}},\
  and\ \bibinfo {author} {\bibfnamefont {T.}~\bibnamefont {Esslinger}},\
  }\bibfield  {title} {\bibinfo {title} {A {Mott} insulator of fermionic atoms
  in an optical lattice},\ }\href {https://doi.org/10.1038/nature07244}
  {\bibfield  {journal} {\bibinfo  {journal} {Nature}\ }\textbf {\bibinfo
  {volume} {455}},\ \bibinfo {pages} {204} (\bibinfo {year}
  {2008})}\BibitemShut {NoStop}%
\bibitem [{\citenamefont {Schneider}\ \emph {et~al.}(2008)\citenamefont
  {Schneider}, \citenamefont {Hackerm{\"u}ller}, \citenamefont {Will},
  \citenamefont {Best}, \citenamefont {Bloch}, \citenamefont {Costi},
  \citenamefont {Helmes}, \citenamefont {Rasch},\ and\ \citenamefont
  {Rosch}}]{Schneider2008}%
  \BibitemOpen
  \bibfield  {author} {\bibinfo {author} {\bibfnamefont {U.}~\bibnamefont
  {Schneider}}, \bibinfo {author} {\bibfnamefont {L.}~\bibnamefont
  {Hackerm{\"u}ller}}, \bibinfo {author} {\bibfnamefont {S.}~\bibnamefont
  {Will}}, \bibinfo {author} {\bibfnamefont {T.}~\bibnamefont {Best}}, \bibinfo
  {author} {\bibfnamefont {I.}~\bibnamefont {Bloch}}, \bibinfo {author}
  {\bibfnamefont {T.~A.}\ \bibnamefont {Costi}}, \bibinfo {author}
  {\bibfnamefont {R.~W.}\ \bibnamefont {Helmes}}, \bibinfo {author}
  {\bibfnamefont {D.}~\bibnamefont {Rasch}},\ and\ \bibinfo {author}
  {\bibfnamefont {A.}~\bibnamefont {Rosch}},\ }\bibfield  {title} {\bibinfo
  {title} {Metallic and insulating phases of repulsively interacting fermions
  in a {3D} optical lattice},\ }\href {https://doi.org/10.1126/science.1165449}
  {\bibfield  {journal} {\bibinfo  {journal} {Science}\ }\textbf {\bibinfo
  {volume} {322}},\ \bibinfo {pages} {1520} (\bibinfo {year}
  {2008})}\BibitemShut {NoStop}%
\bibitem [{\citenamefont {Becker}\ \emph {et~al.}(2010)\citenamefont {Becker},
  \citenamefont {Soltan-Panahi}, \citenamefont {Kronjäger}, \citenamefont
  {Dörscher}, \citenamefont {Bongs},\ and\ \citenamefont
  {Sengstock}}]{Becker2010}%
  \BibitemOpen
  \bibfield  {author} {\bibinfo {author} {\bibfnamefont {C.}~\bibnamefont
  {Becker}}, \bibinfo {author} {\bibfnamefont {P.}~\bibnamefont
  {Soltan-Panahi}}, \bibinfo {author} {\bibfnamefont {J.}~\bibnamefont
  {Kronjäger}}, \bibinfo {author} {\bibfnamefont {S.}~\bibnamefont
  {Dörscher}}, \bibinfo {author} {\bibfnamefont {K.}~\bibnamefont {Bongs}},\
  and\ \bibinfo {author} {\bibfnamefont {K.}~\bibnamefont {Sengstock}},\
  }\bibfield  {title} {\bibinfo {title} {Ultracold quantum gases in triangular
  optical lattices},\ }\href {https://doi.org/10.1088/1367-2630/12/6/065025}
  {\bibfield  {journal} {\bibinfo  {journal} {New J. Phys.}\ }\textbf {\bibinfo
  {volume} {12}},\ \bibinfo {pages} {065025} (\bibinfo {year}
  {2010})}\BibitemShut {NoStop}%
\bibitem [{\citenamefont {Struck}\ \emph {et~al.}(2011)\citenamefont {Struck},
  \citenamefont {Ölschläger}, \citenamefont {Targat}, \citenamefont
  {Soltan-Panahi}, \citenamefont {Eckardt}, \citenamefont {Lewenstein},
  \citenamefont {Windpassinger},\ and\ \citenamefont {Sengstock}}]{Struck2011}%
  \BibitemOpen
  \bibfield  {author} {\bibinfo {author} {\bibfnamefont {J.}~\bibnamefont
  {Struck}}, \bibinfo {author} {\bibfnamefont {C.}~\bibnamefont
  {Ölschläger}}, \bibinfo {author} {\bibfnamefont {R.~L.}\ \bibnamefont
  {Targat}}, \bibinfo {author} {\bibfnamefont {P.}~\bibnamefont
  {Soltan-Panahi}}, \bibinfo {author} {\bibfnamefont {A.}~\bibnamefont
  {Eckardt}}, \bibinfo {author} {\bibfnamefont {M.}~\bibnamefont {Lewenstein}},
  \bibinfo {author} {\bibfnamefont {P.}~\bibnamefont {Windpassinger}},\ and\
  \bibinfo {author} {\bibfnamefont {K.}~\bibnamefont {Sengstock}},\ }\bibfield
  {title} {\bibinfo {title} {Quantum simulation of frustrated classical
  magnetism in triangular optical lattices},\ }\href
  {https://doi.org/10.1126/science.1207239} {\bibfield  {journal} {\bibinfo
  {journal} {Science}\ }\textbf {\bibinfo {volume} {333}},\ \bibinfo {pages}
  {996} (\bibinfo {year} {2011})}\BibitemShut {NoStop}%
\bibitem [{\citenamefont {Jo}\ \emph {et~al.}(2012)\citenamefont {Jo},
  \citenamefont {Guzman}, \citenamefont {Thomas}, \citenamefont {Hosur},
  \citenamefont {Vishwanath},\ and\ \citenamefont {Stamper-Kurn}}]{Jo2012}%
  \BibitemOpen
  \bibfield  {author} {\bibinfo {author} {\bibfnamefont {G.-B.}\ \bibnamefont
  {Jo}}, \bibinfo {author} {\bibfnamefont {J.}~\bibnamefont {Guzman}}, \bibinfo
  {author} {\bibfnamefont {C.~K.}\ \bibnamefont {Thomas}}, \bibinfo {author}
  {\bibfnamefont {P.}~\bibnamefont {Hosur}}, \bibinfo {author} {\bibfnamefont
  {A.}~\bibnamefont {Vishwanath}},\ and\ \bibinfo {author} {\bibfnamefont
  {D.~M.}\ \bibnamefont {Stamper-Kurn}},\ }\bibfield  {title} {\bibinfo {title}
  {Ultracold atoms in a tunable optical kagome lattice},\ }\href
  {https://doi.org/10.1103/PhysRevLett.108.045305} {\bibfield  {journal}
  {\bibinfo  {journal} {Phys. Rev. Lett.}\ }\textbf {\bibinfo {volume} {108}},\
  \bibinfo {pages} {045305} (\bibinfo {year} {2012})}\BibitemShut {NoStop}%
\bibitem [{\citenamefont {Taie}\ \emph {et~al.}(2015)\citenamefont {Taie},
  \citenamefont {Ozawa}, \citenamefont {Ichinose}, \citenamefont {Nishio},
  \citenamefont {Nakajima},\ and\ \citenamefont {Takahashi}}]{Taie2015}%
  \BibitemOpen
  \bibfield  {author} {\bibinfo {author} {\bibfnamefont {S.}~\bibnamefont
  {Taie}}, \bibinfo {author} {\bibfnamefont {H.}~\bibnamefont {Ozawa}},
  \bibinfo {author} {\bibfnamefont {T.}~\bibnamefont {Ichinose}}, \bibinfo
  {author} {\bibfnamefont {T.}~\bibnamefont {Nishio}}, \bibinfo {author}
  {\bibfnamefont {S.}~\bibnamefont {Nakajima}},\ and\ \bibinfo {author}
  {\bibfnamefont {Y.}~\bibnamefont {Takahashi}},\ }\bibfield  {title} {\bibinfo
  {title} {Coherent driving and freezing of bosonic matter wave in an optical
  {Lieb} lattice},\ }\href {https://doi.org/10.1126/sciadv.1500854} {\bibfield
  {journal} {\bibinfo  {journal} {Sci. Adv.}\ }\textbf {\bibinfo {volume}
  {1}},\ \bibinfo {pages} {e1500854} (\bibinfo {year} {2015})}\BibitemShut
  {NoStop}%
\bibitem [{\citenamefont {Yamamoto}\ \emph {et~al.}(2020)\citenamefont
  {Yamamoto}, \citenamefont {Ozawa}, \citenamefont {Nak}, \citenamefont
  {Nakamura},\ and\ \citenamefont {Fukuhara}}]{Yamamoto2020}%
  \BibitemOpen
  \bibfield  {author} {\bibinfo {author} {\bibfnamefont {R.}~\bibnamefont
  {Yamamoto}}, \bibinfo {author} {\bibfnamefont {H.}~\bibnamefont {Ozawa}},
  \bibinfo {author} {\bibfnamefont {D.~C.}\ \bibnamefont {Nak}}, \bibinfo
  {author} {\bibfnamefont {I.}~\bibnamefont {Nakamura}},\ and\ \bibinfo
  {author} {\bibfnamefont {T.}~\bibnamefont {Fukuhara}},\ }\bibfield  {title}
  {\bibinfo {title} {Single-site-resolved imaging of ultracold atoms in a
  triangular optical lattice},\ }\href
  {https://doi.org/10.1088/1367-2630/abcdc8} {\bibfield  {journal} {\bibinfo
  {journal} {New J. Phys.}\ }\textbf {\bibinfo {volume} {22}},\ \bibinfo
  {pages} {123028} (\bibinfo {year} {2020})}\BibitemShut {NoStop}%
\bibitem [{\citenamefont {Tieleman}\ \emph {et~al.}(2013)\citenamefont
  {Tieleman}, \citenamefont {Dutta}, \citenamefont {Lewenstein},\ and\
  \citenamefont {Eckardt}}]{Tieleman2013}%
  \BibitemOpen
  \bibfield  {author} {\bibinfo {author} {\bibfnamefont {O.}~\bibnamefont
  {Tieleman}}, \bibinfo {author} {\bibfnamefont {O.}~\bibnamefont {Dutta}},
  \bibinfo {author} {\bibfnamefont {M.}~\bibnamefont {Lewenstein}},\ and\
  \bibinfo {author} {\bibfnamefont {A.}~\bibnamefont {Eckardt}},\ }\bibfield
  {title} {\bibinfo {title} {Spontaneous time-reversal symmetry breaking for
  spinless fermions on a triangular lattice},\ }\href
  {https://doi.org/10.1103/PhysRevLett.110.096405} {\bibfield  {journal}
  {\bibinfo  {journal} {Phys. Rev. Lett.}\ }\textbf {\bibinfo {volume} {110}},\
  \bibinfo {pages} {096405} (\bibinfo {year} {2013})}\BibitemShut {NoStop}%
\bibitem [{\citenamefont {Cheuk}\ \emph {et~al.}(2015)\citenamefont {Cheuk},
  \citenamefont {Nichols}, \citenamefont {Okan}, \citenamefont {Gersdorf},
  \citenamefont {Ramasesh}, \citenamefont {Bakr}, \citenamefont {Lompe},\ and\
  \citenamefont {Zwierlein}}]{Cheuk2015}%
  \BibitemOpen
  \bibfield  {author} {\bibinfo {author} {\bibfnamefont {L.~W.}\ \bibnamefont
  {Cheuk}}, \bibinfo {author} {\bibfnamefont {M.~A.}\ \bibnamefont {Nichols}},
  \bibinfo {author} {\bibfnamefont {M.}~\bibnamefont {Okan}}, \bibinfo {author}
  {\bibfnamefont {T.}~\bibnamefont {Gersdorf}}, \bibinfo {author}
  {\bibfnamefont {V.~V.}\ \bibnamefont {Ramasesh}}, \bibinfo {author}
  {\bibfnamefont {W.~S.}\ \bibnamefont {Bakr}}, \bibinfo {author}
  {\bibfnamefont {T.}~\bibnamefont {Lompe}},\ and\ \bibinfo {author}
  {\bibfnamefont {M.~W.}\ \bibnamefont {Zwierlein}},\ }\bibfield  {title}
  {\bibinfo {title} {Quantum-gas microscope for fermionic atoms},\ }\href
  {https://doi.org/10.1103/PhysRevLett.114.193001} {\bibfield  {journal}
  {\bibinfo  {journal} {Phys. Rev. Lett.}\ }\textbf {\bibinfo {volume} {114}},\
  \bibinfo {pages} {193001} (\bibinfo {year} {2015})}\BibitemShut {NoStop}%
\bibitem [{\citenamefont {Parsons}\ \emph {et~al.}(2015)\citenamefont
  {Parsons}, \citenamefont {Huber}, \citenamefont {Mazurenko}, \citenamefont
  {Chiu}, \citenamefont {Setiawan}, \citenamefont {Wooley-Brown}, \citenamefont
  {Blatt},\ and\ \citenamefont {Greiner}}]{Parsons2015}%
  \BibitemOpen
  \bibfield  {author} {\bibinfo {author} {\bibfnamefont {M.~F.}\ \bibnamefont
  {Parsons}}, \bibinfo {author} {\bibfnamefont {F.}~\bibnamefont {Huber}},
  \bibinfo {author} {\bibfnamefont {A.}~\bibnamefont {Mazurenko}}, \bibinfo
  {author} {\bibfnamefont {C.~S.}\ \bibnamefont {Chiu}}, \bibinfo {author}
  {\bibfnamefont {W.}~\bibnamefont {Setiawan}}, \bibinfo {author}
  {\bibfnamefont {K.}~\bibnamefont {Wooley-Brown}}, \bibinfo {author}
  {\bibfnamefont {S.}~\bibnamefont {Blatt}},\ and\ \bibinfo {author}
  {\bibfnamefont {M.}~\bibnamefont {Greiner}},\ }\bibfield  {title} {\bibinfo
  {title} {Site-resolved imaging of fermionic $^{6}\mathrm{Li}$ in an optical
  lattice},\ }\href {https://doi.org/10.1103/PhysRevLett.114.213002} {\bibfield
   {journal} {\bibinfo  {journal} {Phys. Rev. Lett.}\ }\textbf {\bibinfo
  {volume} {114}},\ \bibinfo {pages} {213002} (\bibinfo {year}
  {2015})}\BibitemShut {NoStop}%
\bibitem [{\citenamefont {Haller}\ \emph {et~al.}(2015)\citenamefont {Haller},
  \citenamefont {Hudson}, \citenamefont {Kelly}, \citenamefont {Cotta},
  \citenamefont {Peaudecerf}, \citenamefont {Bruce},\ and\ \citenamefont
  {Kuhr}}]{Haller2015}%
  \BibitemOpen
  \bibfield  {author} {\bibinfo {author} {\bibfnamefont {E.}~\bibnamefont
  {Haller}}, \bibinfo {author} {\bibfnamefont {J.}~\bibnamefont {Hudson}},
  \bibinfo {author} {\bibfnamefont {A.}~\bibnamefont {Kelly}}, \bibinfo
  {author} {\bibfnamefont {D.~A.}\ \bibnamefont {Cotta}}, \bibinfo {author}
  {\bibfnamefont {B.}~\bibnamefont {Peaudecerf}}, \bibinfo {author}
  {\bibfnamefont {G.~D.}\ \bibnamefont {Bruce}},\ and\ \bibinfo {author}
  {\bibfnamefont {S.}~\bibnamefont {Kuhr}},\ }\bibfield  {title} {\bibinfo
  {title} {Single-atom imaging of fermions in a quantum-gas microscope},\
  }\href {https://doi.org/10.1038/nphys3403} {\bibfield  {journal} {\bibinfo
  {journal} {Nat. Phys.}\ }\textbf {\bibinfo {volume} {11}},\ \bibinfo {pages}
  {738} (\bibinfo {year} {2015})}\BibitemShut {NoStop}%
\bibitem [{\citenamefont {Edge}\ \emph {et~al.}(2015)\citenamefont {Edge},
  \citenamefont {Anderson}, \citenamefont {Jervis}, \citenamefont {McKay},
  \citenamefont {Day}, \citenamefont {Trotzky},\ and\ \citenamefont
  {Thywissen}}]{Edge2015}%
  \BibitemOpen
  \bibfield  {author} {\bibinfo {author} {\bibfnamefont {G.~J.~A.}\
  \bibnamefont {Edge}}, \bibinfo {author} {\bibfnamefont {R.}~\bibnamefont
  {Anderson}}, \bibinfo {author} {\bibfnamefont {D.}~\bibnamefont {Jervis}},
  \bibinfo {author} {\bibfnamefont {D.~C.}\ \bibnamefont {McKay}}, \bibinfo
  {author} {\bibfnamefont {R.}~\bibnamefont {Day}}, \bibinfo {author}
  {\bibfnamefont {S.}~\bibnamefont {Trotzky}},\ and\ \bibinfo {author}
  {\bibfnamefont {J.~H.}\ \bibnamefont {Thywissen}},\ }\bibfield  {title}
  {\bibinfo {title} {Imaging and addressing of individual fermionic atoms in an
  optical lattice},\ }\href {https://doi.org/10.1103/PhysRevA.92.063406}
  {\bibfield  {journal} {\bibinfo  {journal} {Phys. Rev. A}\ }\textbf {\bibinfo
  {volume} {92}},\ \bibinfo {pages} {063406} (\bibinfo {year}
  {2015})}\BibitemShut {NoStop}%
\bibitem [{\citenamefont {Omran}\ \emph {et~al.}(2015)\citenamefont {Omran},
  \citenamefont {Boll}, \citenamefont {Hilker}, \citenamefont {Kleinlein},
  \citenamefont {Salomon}, \citenamefont {Bloch},\ and\ \citenamefont
  {Gross}}]{Omran2015}%
  \BibitemOpen
  \bibfield  {author} {\bibinfo {author} {\bibfnamefont {A.}~\bibnamefont
  {Omran}}, \bibinfo {author} {\bibfnamefont {M.}~\bibnamefont {Boll}},
  \bibinfo {author} {\bibfnamefont {T.~A.}\ \bibnamefont {Hilker}}, \bibinfo
  {author} {\bibfnamefont {K.}~\bibnamefont {Kleinlein}}, \bibinfo {author}
  {\bibfnamefont {G.}~\bibnamefont {Salomon}}, \bibinfo {author} {\bibfnamefont
  {I.}~\bibnamefont {Bloch}},\ and\ \bibinfo {author} {\bibfnamefont
  {C.}~\bibnamefont {Gross}},\ }\bibfield  {title} {\bibinfo {title}
  {Microscopic observation of {Pauli} blocking in degenerate fermionic lattice
  gases},\ }\href {https://doi.org/10.1103/PhysRevLett.115.263001} {\bibfield
  {journal} {\bibinfo  {journal} {Phys. Rev. Lett.}\ }\textbf {\bibinfo
  {volume} {115}},\ \bibinfo {pages} {263001} (\bibinfo {year}
  {2015})}\BibitemShut {NoStop}%
\bibitem [{\citenamefont {Greif}\ \emph {et~al.}(2016)\citenamefont {Greif},
  \citenamefont {Parsons}, \citenamefont {Mazurenko}, \citenamefont {Chiu},
  \citenamefont {Blatt}, \citenamefont {Huber}, \citenamefont {Ji},\ and\
  \citenamefont {Greiner}}]{Greif2016}%
  \BibitemOpen
  \bibfield  {author} {\bibinfo {author} {\bibfnamefont {D.}~\bibnamefont
  {Greif}}, \bibinfo {author} {\bibfnamefont {M.~F.}\ \bibnamefont {Parsons}},
  \bibinfo {author} {\bibfnamefont {A.}~\bibnamefont {Mazurenko}}, \bibinfo
  {author} {\bibfnamefont {C.~S.}\ \bibnamefont {Chiu}}, \bibinfo {author}
  {\bibfnamefont {S.}~\bibnamefont {Blatt}}, \bibinfo {author} {\bibfnamefont
  {F.}~\bibnamefont {Huber}}, \bibinfo {author} {\bibfnamefont
  {G.}~\bibnamefont {Ji}},\ and\ \bibinfo {author} {\bibfnamefont
  {M.}~\bibnamefont {Greiner}},\ }\bibfield  {title} {\bibinfo {title}
  {Site-resolved imaging of a fermionic {Mott} insulator},\ }\href
  {https://doi.org/10.1126/science.aad9041} {\bibfield  {journal} {\bibinfo
  {journal} {Science}\ }\textbf {\bibinfo {volume} {351}},\ \bibinfo {pages}
  {953} (\bibinfo {year} {2016})}\BibitemShut {NoStop}%
\bibitem [{\citenamefont {Bartenstein}\ \emph {et~al.}(2005)\citenamefont
  {Bartenstein}, \citenamefont {Altmeyer}, \citenamefont {Riedl}, \citenamefont
  {Geursen}, \citenamefont {Jochim}, \citenamefont {Chin}, \citenamefont
  {Denschlag}, \citenamefont {Grimm}, \citenamefont {Simoni}, \citenamefont
  {Tiesinga}, \citenamefont {Williams},\ and\ \citenamefont
  {Julienne}}]{Bartenstein2005}%
  \BibitemOpen
  \bibfield  {author} {\bibinfo {author} {\bibfnamefont {M.}~\bibnamefont
  {Bartenstein}}, \bibinfo {author} {\bibfnamefont {A.}~\bibnamefont
  {Altmeyer}}, \bibinfo {author} {\bibfnamefont {S.}~\bibnamefont {Riedl}},
  \bibinfo {author} {\bibfnamefont {R.}~\bibnamefont {Geursen}}, \bibinfo
  {author} {\bibfnamefont {S.}~\bibnamefont {Jochim}}, \bibinfo {author}
  {\bibfnamefont {C.}~\bibnamefont {Chin}}, \bibinfo {author} {\bibfnamefont
  {J.~H.}\ \bibnamefont {Denschlag}}, \bibinfo {author} {\bibfnamefont
  {R.}~\bibnamefont {Grimm}}, \bibinfo {author} {\bibfnamefont
  {A.}~\bibnamefont {Simoni}}, \bibinfo {author} {\bibfnamefont
  {E.}~\bibnamefont {Tiesinga}}, \bibinfo {author} {\bibfnamefont {C.~J.}\
  \bibnamefont {Williams}},\ and\ \bibinfo {author} {\bibfnamefont {P.~S.}\
  \bibnamefont {Julienne}},\ }\bibfield  {title} {\bibinfo {title} {Precise
  determination of {$^6$Li} cold collision parameters by radio-frequency
  spectroscopy on weakly bound molecules.},\ }\href
  {https://doi.org/10.1103/PhysRevLett.94.103201} {\bibfield  {journal}
  {\bibinfo  {journal} {Phys. Rev. Lett.}\ }\textbf {\bibinfo {volume} {94}},\
  \bibinfo {pages} {103201} (\bibinfo {year} {2005})}\BibitemShut {NoStop}%
\bibitem [{\citenamefont {Z\"urn}\ \emph {et~al.}(2013)\citenamefont {Z\"urn},
  \citenamefont {Lompe}, \citenamefont {Wenz}, \citenamefont {Jochim},
  \citenamefont {Julienne},\ and\ \citenamefont {Hutson}}]{Zuern2013}%
  \BibitemOpen
  \bibfield  {author} {\bibinfo {author} {\bibfnamefont {G.}~\bibnamefont
  {Z\"urn}}, \bibinfo {author} {\bibfnamefont {T.}~\bibnamefont {Lompe}},
  \bibinfo {author} {\bibfnamefont {A.~N.}\ \bibnamefont {Wenz}}, \bibinfo
  {author} {\bibfnamefont {S.}~\bibnamefont {Jochim}}, \bibinfo {author}
  {\bibfnamefont {P.~S.}\ \bibnamefont {Julienne}},\ and\ \bibinfo {author}
  {\bibfnamefont {J.~M.}\ \bibnamefont {Hutson}},\ }\bibfield  {title}
  {\bibinfo {title} {Precise characterization of {$^{6}\mathrm{Li}$} {Feshbach}
  resonances using trap-sideband-resolved {RF} spectroscopy of weakly bound
  molecules},\ }\href {https://doi.org/10.1103/PhysRevLett.110.135301}
  {\bibfield  {journal} {\bibinfo  {journal} {Phys. Rev. Lett.}\ }\textbf
  {\bibinfo {volume} {110}},\ \bibinfo {pages} {135301} (\bibinfo {year}
  {2013})}\BibitemShut {NoStop}%
\bibitem [{\citenamefont {Li}\ \emph {et~al.}(2012)\citenamefont {Li},
  \citenamefont {Corcovilos}, \citenamefont {Wang},\ and\ \citenamefont
  {Weiss}}]{Li2012}%
  \BibitemOpen
  \bibfield  {author} {\bibinfo {author} {\bibfnamefont {X.}~\bibnamefont
  {Li}}, \bibinfo {author} {\bibfnamefont {T.~A.}\ \bibnamefont {Corcovilos}},
  \bibinfo {author} {\bibfnamefont {Y.}~\bibnamefont {Wang}},\ and\ \bibinfo
  {author} {\bibfnamefont {D.~S.}\ \bibnamefont {Weiss}},\ }\bibfield  {title}
  {\bibinfo {title} {{3D} projection sideband cooling},\ }\href
  {https://doi.org/10.1103/PhysRevLett.108.103001} {\bibfield  {journal}
  {\bibinfo  {journal} {Phys. Rev. Lett.}\ }\textbf {\bibinfo {volume} {108}},\
  \bibinfo {pages} {103001} (\bibinfo {year} {2012})}\BibitemShut {NoStop}%
\bibitem [{\citenamefont {Sherson}\ \emph {et~al.}(2010)\citenamefont
  {Sherson}, \citenamefont {Weitenberg}, \citenamefont {Endres}, \citenamefont
  {Cheneau}, \citenamefont {Bloch},\ and\ \citenamefont {Kuhr}}]{Sherson2010}%
  \BibitemOpen
  \bibfield  {author} {\bibinfo {author} {\bibfnamefont {J.~F.}\ \bibnamefont
  {Sherson}}, \bibinfo {author} {\bibfnamefont {C.}~\bibnamefont {Weitenberg}},
  \bibinfo {author} {\bibfnamefont {M.}~\bibnamefont {Endres}}, \bibinfo
  {author} {\bibfnamefont {M.}~\bibnamefont {Cheneau}}, \bibinfo {author}
  {\bibfnamefont {I.}~\bibnamefont {Bloch}},\ and\ \bibinfo {author}
  {\bibfnamefont {S.}~\bibnamefont {Kuhr}},\ }\bibfield  {title} {\bibinfo
  {title} {Single-atom-resolved fluorescence imaging of an atomic {Mott}
  insulator.},\ }\href {https://doi.org/10.1038/nature09378} {\bibfield
  {journal} {\bibinfo  {journal} {Nature}\ }\textbf {\bibinfo {volume} {467}},\
  \bibinfo {pages} {68} (\bibinfo {year} {2010})}\BibitemShut {NoStop}%
\bibitem [{\citenamefont {Anderson}\ \emph {et~al.}(1994)\citenamefont
  {Anderson}, \citenamefont {Petrich}, \citenamefont {Ensher},\ and\
  \citenamefont {Cornell}}]{Anderson1994}%
  \BibitemOpen
  \bibfield  {author} {\bibinfo {author} {\bibfnamefont {M.~H.}\ \bibnamefont
  {Anderson}}, \bibinfo {author} {\bibfnamefont {W.}~\bibnamefont {Petrich}},
  \bibinfo {author} {\bibfnamefont {J.~R.}\ \bibnamefont {Ensher}},\ and\
  \bibinfo {author} {\bibfnamefont {E.~A.}\ \bibnamefont {Cornell}},\
  }\bibfield  {title} {\bibinfo {title} {Reduction of light-assisted
  collisional loss rate from a low-pressure vapor-cell trap},\ }\href
  {https://doi.org/10.1103/PhysRevA.50.R3597} {\bibfield  {journal} {\bibinfo
  {journal} {Phys. Rev. A}\ }\textbf {\bibinfo {volume} {50}},\ \bibinfo
  {pages} {R3597} (\bibinfo {year} {1994})}\BibitemShut {NoStop}%
\bibitem [{\citenamefont {Fuhrmanek}\ \emph {et~al.}(2012)\citenamefont
  {Fuhrmanek}, \citenamefont {Bourgain}, \citenamefont {Sortais},\ and\
  \citenamefont {Browaeys}}]{Fuhrmanek2012}%
  \BibitemOpen
  \bibfield  {author} {\bibinfo {author} {\bibfnamefont {A.}~\bibnamefont
  {Fuhrmanek}}, \bibinfo {author} {\bibfnamefont {R.}~\bibnamefont {Bourgain}},
  \bibinfo {author} {\bibfnamefont {Y.~R.~P.}\ \bibnamefont {Sortais}},\ and\
  \bibinfo {author} {\bibfnamefont {A.}~\bibnamefont {Browaeys}},\ }\bibfield
  {title} {\bibinfo {title} {Light-assisted collisions between a few cold atoms
  in a microscopic dipole trap},\ }\href
  {https://doi.org/10.1103/PhysRevA.85.062708} {\bibfield  {journal} {\bibinfo
  {journal} {Phys. Rev. A}\ }\textbf {\bibinfo {volume} {85}},\ \bibinfo
  {pages} {062708} (\bibinfo {year} {2012})}\BibitemShut {NoStop}%
\bibitem [{\citenamefont {Endres}\ \emph {et~al.}(2013)\citenamefont {Endres},
  \citenamefont {Cheneau}, \citenamefont {Fukuhara}, \citenamefont
  {Weitenberg}, \citenamefont {Schau{\ss}}, \citenamefont {Gross},
  \citenamefont {Mazza}, \citenamefont {Ba{\~{n}}uls}, \citenamefont {Pollet},
  \citenamefont {Bloch},\ and\ \citenamefont {Kuhr}}]{Endres2013}%
  \BibitemOpen
  \bibfield  {author} {\bibinfo {author} {\bibfnamefont {M.}~\bibnamefont
  {Endres}}, \bibinfo {author} {\bibfnamefont {M.}~\bibnamefont {Cheneau}},
  \bibinfo {author} {\bibfnamefont {T.}~\bibnamefont {Fukuhara}}, \bibinfo
  {author} {\bibfnamefont {C.}~\bibnamefont {Weitenberg}}, \bibinfo {author}
  {\bibfnamefont {P.}~\bibnamefont {Schau{\ss}}}, \bibinfo {author}
  {\bibfnamefont {C.}~\bibnamefont {Gross}}, \bibinfo {author} {\bibfnamefont
  {L.}~\bibnamefont {Mazza}}, \bibinfo {author} {\bibfnamefont {M.~C.}\
  \bibnamefont {Ba{\~{n}}uls}}, \bibinfo {author} {\bibfnamefont
  {L.}~\bibnamefont {Pollet}}, \bibinfo {author} {\bibfnamefont
  {I.}~\bibnamefont {Bloch}},\ and\ \bibinfo {author} {\bibfnamefont
  {S.}~\bibnamefont {Kuhr}},\ }\bibfield  {title} {\bibinfo {title}
  {Single-site- and single-atom-resolved measurement of correlation
  functions},\ }\href {https://doi.org/10.1007/s00340-013-5552-9} {\bibfield
  {journal} {\bibinfo  {journal} {Appl. Phys. B}\ }\textbf {\bibinfo {volume}
  {113}},\ \bibinfo {pages} {27} (\bibinfo {year} {2013})}\BibitemShut
  {NoStop}%
\bibitem [{\citenamefont {Sebby-Strabley}\ \emph {et~al.}(2006)\citenamefont
  {Sebby-Strabley}, \citenamefont {Anderlini}, \citenamefont {Jessen},\ and\
  \citenamefont {Porto}}]{Sebby2006}%
  \BibitemOpen
  \bibfield  {author} {\bibinfo {author} {\bibfnamefont {J.}~\bibnamefont
  {Sebby-Strabley}}, \bibinfo {author} {\bibfnamefont {M.}~\bibnamefont
  {Anderlini}}, \bibinfo {author} {\bibfnamefont {P.~S.}\ \bibnamefont
  {Jessen}},\ and\ \bibinfo {author} {\bibfnamefont {J.~V.}\ \bibnamefont
  {Porto}},\ }\bibfield  {title} {\bibinfo {title} {Lattice of double wells for
  manipulating pairs of cold atoms},\ }\href
  {https://doi.org/10.1103/PhysRevA.73.033605} {\bibfield  {journal} {\bibinfo
  {journal} {Phys. Rev. A}\ }\textbf {\bibinfo {volume} {73}},\ \bibinfo
  {pages} {033605} (\bibinfo {year} {2006})}\BibitemShut {NoStop}%
\bibitem [{\citenamefont {Sbroscia}\ \emph {et~al.}(2020)\citenamefont
  {Sbroscia}, \citenamefont {Viebahn}, \citenamefont {Carter}, \citenamefont
  {Yu}, \citenamefont {Gaunt},\ and\ \citenamefont {Schneider}}]{Sbroscia2020}%
  \BibitemOpen
  \bibfield  {author} {\bibinfo {author} {\bibfnamefont {M.}~\bibnamefont
  {Sbroscia}}, \bibinfo {author} {\bibfnamefont {K.}~\bibnamefont {Viebahn}},
  \bibinfo {author} {\bibfnamefont {E.}~\bibnamefont {Carter}}, \bibinfo
  {author} {\bibfnamefont {J.-C.}\ \bibnamefont {Yu}}, \bibinfo {author}
  {\bibfnamefont {A.}~\bibnamefont {Gaunt}},\ and\ \bibinfo {author}
  {\bibfnamefont {U.}~\bibnamefont {Schneider}},\ }\bibfield  {title} {\bibinfo
  {title} {Observing localization in a {2D} quasicrystalline optical lattice},\
  }\href {https://doi.org/10.1103/PhysRevLett.125.200604} {\bibfield  {journal}
  {\bibinfo  {journal} {Phys. Rev. Lett.}\ }\textbf {\bibinfo {volume} {125}},\
  \bibinfo {pages} {200604} (\bibinfo {year} {2020})}\BibitemShut {NoStop}%
\bibitem [{\citenamefont {Brown}\ \emph {et~al.}(2019)\citenamefont {Brown},
  \citenamefont {Mitra}, \citenamefont {Guardado-Sanchez}, \citenamefont
  {Nourafkan}, \citenamefont {Reymbaut}, \citenamefont {Hébert}, \citenamefont
  {Bergeron}, \citenamefont {Tremblay}, \citenamefont {Kokalj}, \citenamefont
  {Huse}, \citenamefont {Schauß},\ and\ \citenamefont {Bakr}}]{Brown2019}%
  \BibitemOpen
  \bibfield  {author} {\bibinfo {author} {\bibfnamefont {P.~T.}\ \bibnamefont
  {Brown}}, \bibinfo {author} {\bibfnamefont {D.}~\bibnamefont {Mitra}},
  \bibinfo {author} {\bibfnamefont {E.}~\bibnamefont {Guardado-Sanchez}},
  \bibinfo {author} {\bibfnamefont {R.}~\bibnamefont {Nourafkan}}, \bibinfo
  {author} {\bibfnamefont {A.}~\bibnamefont {Reymbaut}}, \bibinfo {author}
  {\bibfnamefont {C.-D.}\ \bibnamefont {Hébert}}, \bibinfo {author}
  {\bibfnamefont {S.}~\bibnamefont {Bergeron}}, \bibinfo {author}
  {\bibfnamefont {A.-M.~S.}\ \bibnamefont {Tremblay}}, \bibinfo {author}
  {\bibfnamefont {J.}~\bibnamefont {Kokalj}}, \bibinfo {author} {\bibfnamefont
  {D.~A.}\ \bibnamefont {Huse}}, \bibinfo {author} {\bibfnamefont
  {P.}~\bibnamefont {Schauß}},\ and\ \bibinfo {author} {\bibfnamefont {W.~S.}\
  \bibnamefont {Bakr}},\ }\bibfield  {title} {\bibinfo {title} {Bad metallic
  transport in a cold atom {Fermi}-{Hubbard} system},\ }\href
  {https://doi.org/10.1126/science.aat4134} {\bibfield  {journal} {\bibinfo
  {journal} {Science}\ }\textbf {\bibinfo {volume} {363}},\ \bibinfo {pages}
  {379} (\bibinfo {year} {2019})}\BibitemShut {NoStop}%
\bibitem [{\citenamefont {Vrani\ifmmode~\acute{c}\else \'{c}\fi{}}\ \emph
  {et~al.}(2020)\citenamefont {Vrani\ifmmode~\acute{c}\else \'{c}\fi{}},
  \citenamefont {Vu\ifmmode \check{c}\else \v{c}\fi{}i\ifmmode \check{c}\else
  \v{c}\fi{}evi\ifmmode~\acute{c}\else \'{c}\fi{}}, \citenamefont {Kokalj},
  \citenamefont {Skolimowski}, \citenamefont {\ifmmode~\check{Z}\else
  \v{Z}\fi{}itko}, \citenamefont {Mravlje},\ and\ \citenamefont
  {Tanaskovi\ifmmode~\acute{c}\else \'{c}\fi{}}}]{Vranic2020}%
  \BibitemOpen
  \bibfield  {author} {\bibinfo {author} {\bibfnamefont {A.}~\bibnamefont
  {Vrani\ifmmode~\acute{c}\else \'{c}\fi{}}}, \bibinfo {author} {\bibfnamefont
  {J.}~\bibnamefont {Vu\ifmmode \check{c}\else \v{c}\fi{}i\ifmmode
  \check{c}\else \v{c}\fi{}evi\ifmmode~\acute{c}\else \'{c}\fi{}}}, \bibinfo
  {author} {\bibfnamefont {J.}~\bibnamefont {Kokalj}}, \bibinfo {author}
  {\bibfnamefont {J.}~\bibnamefont {Skolimowski}}, \bibinfo {author}
  {\bibfnamefont {R.}~\bibnamefont {\ifmmode~\check{Z}\else \v{Z}\fi{}itko}},
  \bibinfo {author} {\bibfnamefont {J.}~\bibnamefont {Mravlje}},\ and\ \bibinfo
  {author} {\bibfnamefont {D.}~\bibnamefont {Tanaskovi\ifmmode~\acute{c}\else
  \'{c}\fi{}}},\ }\bibfield  {title} {\bibinfo {title} {Charge transport in the
  {Hubbard} model at high temperatures: {Triangular} versus square lattice},\
  }\href {https://doi.org/10.1103/PhysRevB.102.115142} {\bibfield  {journal}
  {\bibinfo  {journal} {Phys. Rev. B}\ }\textbf {\bibinfo {volume} {102}},\
  \bibinfo {pages} {115142} (\bibinfo {year} {2020})}\BibitemShut {NoStop}%
\bibitem [{\citenamefont {Zhang}\ \emph {et~al.}(2018)\citenamefont {Zhang},
  \citenamefont {Zhu},\ and\ \citenamefont {Batista}}]{Zhang2018}%
  \BibitemOpen
  \bibfield  {author} {\bibinfo {author} {\bibfnamefont {S.-S.}\ \bibnamefont
  {Zhang}}, \bibinfo {author} {\bibfnamefont {W.}~\bibnamefont {Zhu}},\ and\
  \bibinfo {author} {\bibfnamefont {C.~D.}\ \bibnamefont {Batista}},\
  }\bibfield  {title} {\bibinfo {title} {Pairing from strong repulsion in
  triangular lattice {Hubbard} model},\ }\href
  {https://doi.org/10.1103/PhysRevB.97.140507} {\bibfield  {journal} {\bibinfo
  {journal} {Phys. Rev. B}\ }\textbf {\bibinfo {volume} {97}},\ \bibinfo
  {pages} {140507} (\bibinfo {year} {2018})}\BibitemShut {NoStop}%
\end{thebibliography}

%

\end{document}